\documentclass[prd,aps,letter,twocolumn,floatfix,notitlepage,nofootinbib]{revtex4-1}

\usepackage{graphicx,psfrag}
\usepackage{mathrsfs,amsmath,amsfonts,amssymb, bm}
\usepackage{multirow}
\usepackage{comment,hyperref}
\usepackage{float}
\usepackage{algorithm}
\usepackage{algpseudocode} 
\usepackage{setspace} 
\usepackage{times}
\usepackage{ulem}

\newcommand{\be}{\begin{equation}}
\newcommand{\ee}{\end{equation}}
\newcommand{\bsube}{\begin{subequations}}
\newcommand{\esube}{\end{subequations}}

\newcommand{\nnm}{\nonumber}

\newcommand{\bea}{\begin{eqnarray}}
\newcommand{\eea}{\end{eqnarray}}
\newcommand{\bdm}{\begin{displaymath}}
\newcommand{\edm}{\end{displaymath}}
\newcommand{\bse}{\begin{subequations}}
\newcommand{\ese}{\end{subequations}}

\newcommand{\mr}{\mathrm}
\newcommand{\tr}{\textrm}
\newcommand{\mc}{\mathcal}

\newcommand{\bs}{\boldsymbol}

\def\bx{\mathbf{x}}
\def\by{\mathbf{y}}
\def\btheta{\boldsymbol{\theta}}

\usepackage{color}
\definecolor{cyan}{rgb}{0,0.9,0.9}
\definecolor{orange}{rgb}{0.9,0.5,0}
\definecolor{magenta}{rgb}{1,0,1}
\definecolor{purple}{rgb}{0.8,0.4,0.8}
\definecolor{gray}{rgb}{0.8242,0.8242,0.8242}

\begin{document}

\title{Surrogate model for an aligned-spin effective one body waveform model of binary neutron star inspirals using Gaussian process regression}

\author{
Benjamin D. Lackey, 
Michael P\"{u}rrer, 
Andrea Taracchini,
Sylvain Marsat
}
\affiliation{
Max Planck Institute for Gravitational Physics, Albert Einstein Institute, D-14476 Golm, Germany
} 

\date{\today}

\begin{abstract}

Fast and accurate waveform models are necessary for measuring the properties of inspiraling binary neutron star systems such as GW170817. We present a frequency-domain surrogate version of the aligned-spin binary neutron star waveform model using the effective one body formalism known as \texttt{SEOBNRv4T}. This model includes the quadrupolar and octopolar adiabatic and dynamical tides. The version presented here is improved by the inclusion of the spin-induced quadrupole moment effect, and completed by a prescription for tapering the end of the waveform to qualitatively reproduce numerical relativity simulations. The resulting model has 14 intrinsic parameters. We reduce its dimensionality by using universal relations that approximate all matter effects in terms of the leading quadrupolar tidal parameters. The implementation of the time-domain model can take up to an hour to evaluate using a starting frequency of 20~Hz, and this is too slow for many parameter estimation codes that require $\mathcal{O}(10^7)$ sequential waveform evaluations. We therefore construct a fast and faithful frequency-domain surrogate of this model using Gaussian process regression. The resulting surrogate has a maximum mismatch of $4.5\times 10^{-4}$ for the Advanced LIGO detector, and requires $\sim 0.13$~s to evaluate for a waveform with a starting frequency of 20~Hz. Finally, we perform an end-to-end test of the surrogate with a set of parameter estimation runs, and find that the surrogate accurately recovers the parameters of injected waveforms. 

\end{abstract}


\maketitle


\section{Introduction}

The detection of the merging binary neutron star (BNS) system GW170817~\cite{GW170817} has demonstrated that the Advanced LIGO~\cite{Harry2010} and Advanced Virgo~\cite{Acernese2009} gravitational-wave detectors can measure BNS properties such as the masses, spins, and tidal parameters~\cite{BNSPE}, and these results can even be extended to measure quantities such as the radii and equation of state (EOS) of NSs~\cite{RaithelOzelPsaltis2018, DeFinstadLattimer2018, EOSPaper}. Furthermore, an estimate of the BNS merger rate of 110--3840 events Gpc$^{-3}$~yr$^{-1}$ (90\% confidence interval)~\cite{O2Catalog} from this event indicates that, when second generation detectors reach design sensitivity, we may eventually observe a population of tens or hundreds of events after a couple of years of observation~\cite{LIGORate2010}, and KAGRA~\cite{Somiya2012} and LIGO-India~\cite{IyerSouradeepUnnikrishnan2011} could improve this further. These multiple events could then be ``stacked'' to significantly improve measurements of the EOS~\cite{DelPozzoLiAgathos2013, LackeyWade2015}, and also measure the population distribution of masses and spins. 

The reliability of these measurements, however, depends on Bayesian analyses that require fast and accurate waveform models to match to the gravitational-wave data. Standard Bayesian parameter estimation tools such as Markov chain Monte Carlo (MCMC) and Nested Sampling often require $\mathcal{O}(10^7)$ sequential waveform evaluations. Waveform evaluation times must therefore be significantly less than 1~s in order to run in less than a month for each event. To avoid large systematic errors, highly accurate waveform models are needed that include all relevant physical effects such as point-particle interactions, spins, and tides~\cite{Favata2014, YagiYunes2014, WadeCreightonOchsner2014}. Most time-domain waveform implementations involve solving a set of ordinary differential equations, which depending on the model can take seconds to hours for long BNS waveforms. The results then need to be transformed into the frequency domain where the analysis takes place. Because of this, analytic approximations providing waveforms directly in the frequency-domain  are often used instead. 

For the analysis of GW170817 in Refs.~\cite{BNSPE, O2Catalog}, therefore, a set of four frequency-domain waveform models were used as templates for the main results. These models are modifications to spinning binary black hole (BBH) waveform models (see Table I of Ref.~\cite{BNSPE} and references therein). They use either an aligned-spin post-Newtonian (PN) approximant in the Fourier domain (TaylorF2), the aligned-spin effective one body (EOB) formalism (SEOBNRT), or the phenomenological formalism for aligned-spin (PhenomDNRT) and precessing spin (PhenomPNRT). Tidal interactions were then included by simply adding a correction to the phase of each waveform. For the TaylorF2 waveform, the analytic 5PN and 6PN order terms were added. For the other three models, a fit that combined results from PN and numerical BNS simulations (referred to as the NRTidal fit in Ref.~\cite{BNSPE}) was added~\cite{DietrichBernuzziTichy2017}.

In this paper, we will use an implementation of the EOB formalism that treats matter effects consistently with the other parameters. It includes several dynamical effects beyond the standard adiabatic inspiral evolution and can also be tuned to numerical relativity (NR) simulations. The model, named \texttt{SEOBNRv4T}, includes the tidally induced $\ell=2$ and 3 multipole moments, the induced $\ell=2$ and 3 $f$-mode resonances~\cite{Hinderer:2016eia,Steinhoff:2016rfi}, and the spin-induced quadrupole moment that can be important for large spins~\cite{Poisson1998, HarryHinderer2018}. This model agrees with NR simulations of BNS systems to a level consistent with the numerical error of the simulations (less than 1~rad during the last $\sim 10$ orbits before merger)~\cite{DietrichHinderer2017, KiuchiKawaguchiKyutoku2017}. It also includes a prescription for tapering the end of the waveform after merger in a manner consistent with numerical simulations. We also note there is an alternative EOB model, \texttt{TEOBResumS}~\cite{NagarBernuzziDelpozzo2018, NagarMessinaRettegno2018}, that uses a different point-particle and spin prescription. For matter effects, this model includes the $\ell = 2, 3$, and 4 tidal terms and uses a method for re-summing these expressions. It also includes the spin induced-quadrupole term, but does not include the effect of dynamical tides from the induced $f$-mode resonances. 

Although these EOB models are more expensive than frequency-domain models, parameter estimation results are still possible. In Refs.~\cite{BNSPE, O2Catalog}, an alternative parallelized code, \texttt{RapidPE}, was also used that approximately fits the posterior density function (PDF) for the intrinsic waveform parameters instead of directly sampling it~\cite{PankowBradyOchsner2015, LangeOshaughnessyRizzo2018}. 
The measured tidal parameter using EOB waveforms and \texttt{RapidPE} were broadly consistent with the results using frequency-domain waveforms and traditional samplers (see Fig. 9 of Ref.~\cite{O2Catalog}). However, a direct comparison using identical waveforms has not yet been published for BNS systems.
As another possible solution for doing parameter estimation with EOB models, significant work has been done to optimize some EOB implementations, and speedups of a factor of several hundred~\cite{DevineEtienneMcWilliams2016} have been achieved. Different formulations of the differential equations can also speed up integration~\cite{NagarNettegno2018}. However, many of these optimizations are specific to each waveform model, and it is not clear that sufficient speed ups can be achieved for the \texttt{SEOBNRv4T} model discussed here with dynamical tides. As an alternative to these methods, we will develop a surrogate model of \texttt{SEOBNRv4T} that bypasses the need for specialized parameter estimation tools and optimization techniques. 

Surrogate modeling techniques have had significant success in gravitational-wave data analysis for rapidly evaluating waveform models. The essential idea is to construct a fitting function that approximates a waveform model as a function of frequency (or time) and waveform parameters $\bx$. Previous works have focused on efficiently representing the space of all waveforms using a reduced basis of orthonormal functions. The two most common approaches for selecting a reduced basis have been singular value decomposition~\cite{Puerrer2014, Puerrer2015} and a greedy method~\cite{FieldGalleyHesthaven2014, LackeyBernuzziGalley2017, BlackmanFieldGalley2015, BlackmanFieldScheel2017a, BlackmanFieldScheel2017b}. In this work, instead of using a reduced basis, we will instead use the fact that the Fourier-transformed EOB waveform is approximately known analytically via the fast, analytic TaylorF2 model. We then build a surrogate of the difference (or residual) between the EOB waveform and the TaylorF2 waveform. Because this residual is small, extremely high accuracy is not required, and we can efficiently represent the residual as a function of frequency using cubic splines between a small set of frequency nodes.

The final step in building a surrogate is to interpolate between waveform parameters $\bx$. Many previous works focused on parameter spaces with three or fewer dimensions. This allows one to approximate the waveform as a function of $\bx$ using standard interpolation techniques such as tensor spline~\cite{Puerrer2014, Puerrer2015} or Chebyshev interpolation~\cite{LackeyBernuzziGalley2017}. These interpolation techniques typically require waveforms to be evaluated on a rectangular grid, and thus suffer from the curse of dimensionality: for $N$ points per dimension $d$,  $N^d$ waveforms are needed. For the EOB waveform here, we will only be able to reduce the parameter space to five dimensions, so grid based interpolation is unfeasible. Recent work on optimally choosing waveforms for analytic~\cite{DoctorFarrHolz2017} and NR~\cite{BlackmanFieldScheel2017a, BlackmanFieldScheel2017b} waveforms has shown that there are sufficiently accurate alternatives to grid-based interpolation that do not suffer from the curse of dimensionality.

In this paper we focus on a technique known as Gaussian process regression (GPR)~\cite{RasmussenWilliams2006} that does not require a regular grid. Importantly, GPR also provides a convenient estimate of its own uncertainty. This allows us to iteratively add new points to the training set in such a way as to minimize the interpolation error over the parameter space for a given number of waveforms in the training set. Similar approaches have recently been used to optimally sample waveforms for a 2-dimensional aligned spin BBH surrogate using GPR~\cite{DoctorFarrHolz2017} as well as for a surrogate of nonspinning, eccentric BBH mergers~\cite{HuertaMooreKumar2017}. GPR has also been used in GW data analysis to marginalize over waveform uncertainties in parameter estimation~\cite{MooreGair2014, MooreBerryChua2016}.

This work also takes a different approach from the surrogate developed in Ref.~\cite{LackeyBernuzziGalley2017} for a nonspinning BNS EOB waveform. Whereas in Ref.~\cite{LackeyBernuzziGalley2017} the authors constructed a time-domain surrogate with the goal of reproducing the original waveform model as accurately as possible, in this work, we construct a frequency-domain model that can be extended down to arbitrarily low frequencies and does not require an online Fourier transform for each waveform evaluation. It also enables additional techniques to accelerate likelihood evaluations such as reduced order quadrature~\cite{Antil2013, CanizaresFieldGair2013, CanizaresFieldGair2015, Smith:2016qas}, multi-banding~\cite{VinciguerraVeitchMandel2017}, and relative binning~\cite{Zackay:2018qdy}.

We organize the paper as follows. In Section~\ref{sec:eob} we provide an overview of the \texttt{SEOBNRv4T} waveform model and the approximations used to reduce the dimensionality of the parameter space. In Section~\ref{sec:surrogate} we describe the details of building a surrogate model, choosing training set waveforms, and evaluating the final model. We then compare the accuracy and speed of the surrogate to the original model in Section~\ref{sec:performance}. In Section~\ref{sec:pe} we verify that the surrogate can correctly extract the parameters of injected waveforms. Finally, we discuss future improvements in Section~\ref{sec:discussion} and give the expressions for the TaylorF2 base model in the Appendix.

\textit{Conventions:} Unless explicitly stated, we use units where $G=c=1$.

\section{Aligned-spin, dynamical tides EOB model}
\label{sec:eob}

\subsection{Inspiral-plunge waveform}

The EOB approach to the general-relativistic 2-body problem, first described in Ref.~\cite{Buonanno:1998gg}, has proven successful in modeling the dynamics and GW emission of compact binaries. State-of-the-art aligned-spin EOB models~\cite{Bohe:2016gbl,Nagar:2017jdw} can accurately match hundreds of NR simulations of aligned-spin BBH systems for mass ratios up to 8 and spin magnitudes up to 0.85 for unequal-mass (up to 0.98 for equal-mass) binaries. The EOB framework can also accommodate precessing-spin BBHs, showing good agreement to mildly precessing NR simulations~\cite{Babak:2016tgq}. 

This research program has been extended to accommodate tidal effects~\cite{Damour:2009wj,Vines:2010ca,Damour:2012yf,Bini:2012gu,Bernuzzi:2014owa,Hinderer:2016eia,Steinhoff:2016rfi,Dietrich:2017feu} for binaries that contain NSs. The main effect of tides is to make the gravitational interaction more attractive with respect to the vacuum case. In particular, Refs.~\cite{Hinderer:2016eia,Steinhoff:2016rfi} built upon the aligned-spin BBH EOB model of Ref.~\cite{Taracchini:2013rva} and proposed a way to include the effect of dynamical tides. Neutron stars that are part of a compact-object binary will deform in the tidal field generated by the companion. The forcing tidal field varies at a multiple of the orbital frequency. Thus, in the late stages of the inspiral, the characteristic $f$-mode frequency of the neutron star can be dynamically approached, resulting in a resonant excitation of the $f$-mode. The net effect is an amplification of tidal effects as compared to the adiabatic limit, which assumes that the $f$-mode frequency is much larger than the frequency of the forcing tidal field.

Dynamical tidal effects are implemented in the EOB model through a modification of the potential $\Delta_u$, which is the $tt$-component of the metric of the effective spacetime. We adopt the tidally-augmented expression for $\Delta_u$ discussed in Appendix~A of Ref.~\cite{Steinhoff:2016rfi}:
\begin{equation}
\Delta_u = \Delta_u^{\textrm{pm}} + \Delta_u^{\textrm{DT}}\,,
\end{equation}
where $\Delta_u^{\textrm{pm}}$ is the 4PN-accurate point-mass EOB term (Eq.~(2.2) of Ref.~\cite{Bohe:2016gbl}), and $\Delta_u^{\textrm{DT}}$ is the contribution due to dynamical tides. If either component of the binary is a black hole, then we set the tidal polarizabilities to zero. The tidal contribution, including quadrupolar and octupolar dynamical tides, is
\begin{widetext}
\begin{align}
 \Delta_u^{\textrm{DT}} =&  - 3\,\Lambda_{2,\textrm{dyn}}^{A}(u)X_A^4 X_B\,u^6  \left[ 1 + \frac{5}{2} X_Au + \left(3 + \frac{1}{8}X_A + \frac{337}{28} X_A^2\right)u^2 \right]\nonumber\\
 &- 15\,\Lambda_{3,\textrm{dyn}}^{A}(u)X_A^6X_B\,u^8 \left[1 + \left(-2 + \frac{15}{2}X_A\right)u + \left(\frac{8}{3} - \frac{311}{24}X_A + \frac{110}{3}X_A^2\right) u^2\right] + (A \leftrightarrow B)\,.
\end{align}
\end{widetext}
Here, $m_{A,B}$ are the masses of bodies $A$ and $B$, $M = m_A + m_B$ is the total mass, $X_{A,B}=m_{A,B}/M$, $u=1/r$ is the inverse of the ($M$-rescaled) EOB radial coordinate $r$, and $\Lambda_{\ell,\textrm{dyn}}^{A,B}(u)$ are the dimensionless $2^{\ell}$-polar dynamical tidal polarizabilities. Within the dynamical tides model, the tidal polarizabilities are not constant, but rather depend on the orbital separation and on the values of the $f$-mode frequencies $\hat{\omega}_{0\ell}^{A,B}$\footnote{Here, $\hat{\omega}_{0\ell}^{A,B}=m_{A,B}\omega_{0\ell}^{A,B}$, where $\omega_{0\ell}^{A,B}$ is the frequency in geometrized units.}. In particular, the dimensionless dynamical tidal polarizability reads
\begin{equation}
\Lambda_{\ell,\textrm{dyn}}^{A,B}(u)=\Lambda_{\ell}^{A,B}\hat{k}_{\ell, \textrm{dyn}}(u;\hat{\omega}_{0\ell}^{A,B})\,,
\end{equation}
where $\Lambda_{\ell}^{A,B}$ is the dimensionless adiabatic tidal polarizability
\begin{equation}
\Lambda_{\ell}^{A,B}=\frac{2}{(2\ell-1)!!}\frac{k^{A,B}_{\ell}}{C_{A,B}^{2\ell+1}}\,,
\end{equation}
with $k^{A,B}_{\ell}$ the tidal Love number and $C_{A,B}=R_{A,B}/m_{A,B}$ the NS compactness, which depends on the NS radius $R_{A,B}$. Here, $\hat{k}_{\ell, \textrm{dyn}}(u;\hat{\omega}_{0\ell}^{A,B})$ is the separation-dependent, dimensionless enhancement factor~\cite{Steinhoff:2016rfi, Dietrich:2017feu}, which depends on the value of the $f$-mode $2^\ell$-pole dimensionless angular frequency, $\hat{\omega}_{0\ell}^{A,B}$.

Introducing a Keplerian orbital frequency defined from the radius variable as $M\Omega \equiv u^{3/2}$, using the notations (dropping the indices $A,B$)
\begin{subequations}
\begin{align}
	x = \frac{\hat{\omega}_{0\ell}}{m\Omega} \,, \; \epsilon \equiv \frac{256\eta}{5} \left( \frac{\hat{\omega}_{0\ell}}{m} \right)^{5/3} \,, \; \hat{t} \equiv \frac{8}{5\sqrt{\epsilon}} \left[ 1 - x^{5/3} \right] \,,
\end{align}
\end{subequations}
the dynamical enhancement factor  of~\cite{Steinhoff:2016rfi, Dietrich:2017feu} reads
\be\label{eq:kldyn}
	\hat{k}_{\ell, \rm{dyn}}(u; \omega_{0\ell}) = a_{\ell} + b_{\ell} \left[ f(x) +\sqrt{\frac{\pi}{3 \epsilon}} x^{2} \mathcal{Q} \right]\,,
\ee
with $a_{\ell}$, $b_{\ell}$ constant coefficients with relevant values $\{a_{2}, b_{2}\} = \{1/4, 3/4\}$ and $\{a_{3}, b_{3}\} = \{3/8, 5/8\}$. These expressions are used in practice for $\ell= m$. Here, we have rewritten the first two resonant terms of Eq.~(11) of~\cite{Dietrich:2017feu} as
\be\label{eq:deff}
	f(x) = x^{2} \left[ \frac{1}{x^{2} - 1} + \frac{5}{6} \frac{1}{1 - x^{5/3}} \right] \,.
\ee
Although the two terms are individually divergent\footnote{This requires to single out the near-resonance region in a numerical implementation.} at the resonance $x=1$, the function $f(x)$ is actually  regular, taking the value $f(1) = -1/12$. The last term in~\eqref{eq:kldyn} above reads
\begin{align}
	\mathcal{Q} &= \cos\left( \frac{3}{8} \hat{t}^{2} \right) \left[ 1 + 2 F_{S} \left( \frac{\sqrt{3}}{2\sqrt{\pi}} \hat{t} \right) \right] \nonumber\\
	& - \sin\left( \frac{3}{8} \hat{t}^{2} \right) \left[ 1 + 2 F_{C} \left( \frac{\sqrt{3}}{2\sqrt{\pi}} \hat{t} \right) \right] \,,
\end{align}
with $F_{S}$, $F_{C}$ the Fresnel sine and cosine functions. This term is regular at resonance $\hat{t} = 0$. The enhancement factor~\eqref{eq:kldyn} is 1 for low orbital frequencies, and increases to $\sim 2$ as $\ell$ times the orbital frequency approaches the $f$-mode frequency.

Tidal effects also enter the radiative part of the model. In particular, the point-mass $\ell=2,3$ inspiral-plunge waveform modes are corrected by tidal terms
\begin{equation}
h_{\ell m}^{\textrm{insp-plunge}} = h_{\ell m}^{\textrm{pm}} + h_{\ell m}^{\textrm{tidal}}\,,\label{hlminspplunge}
\end{equation} 
where the point-mass piece $h_{\ell m}^{\textrm{pm}}$ is discussed in Section~II.B of Ref.~\cite{Bohe:2016gbl} and the tidal piece $h_{\ell m}^{\textrm{tidal}}$ is given by Eqs. (A14)-(A17) of Ref.~\cite{Damour:2012yf}. Following Ref.~\cite{Dietrich:2017feu} in the computation of $h_{22}^{\textrm{tidal}}$, we include a dynamical enhancement factor that depends on the orbital separation and on the $f$-mode frequency, and is given in Eq.~(15) of Ref.~\cite{Dietrich:2017feu}.

The waveform modes $h_{\ell m}^{\textrm{insp-plunge}}$ are then used to calculate the gravitational-wave flux (Eqs.~(5) and~(6) of Ref.~\cite{Dietrich:2017feu}) from which the radiation-reaction force is derived. The Hamiltonian equations of motion with this radiation-reaction force are numerically integrated beginning with the quasicircular initial conditions described in Ref.~\cite{Buonanno:2005xu}. The inspiral-plunge waveform is obtained by evaluating Eq.~(\ref{hlminspplunge}) using the solution to the orbital dynamics. We only include the dominant $(\ell,m)=(2,2)$ mode in constructing the final waveform, although we use all modes when evaluating the radiation reaction force.

\subsection{Spin-induced quadrupole-monopole terms}
\label{subsec:QM}

For the model used in this paper, \texttt{SEOBNRv4T}, spin effects are based on the point-mass model \texttt{SEOBNRv4}~\cite{Bohe:2016gbl} that was calibrated to 141 NR simulations of BBH systems with aligned spins. It includes consistently spin-orbit interactions up to 3.5PN order and spin-spin interactions up to 2PN order. In addition, the \texttt{SEOBNRv4T} model accounts for the spin-induced quadrupole moment of neutron stars~\cite{Poisson1998}, an extended-body effect distinguishing neutron stars from black holes. This effect is quadratic in the spins, appears at 2PN order as compared to the tidal effects that first appear at 5PN order, and is significant for systems with large spin~\cite{HarryHinderer2018}. Although complete PN expressions for this contribution in the dynamics and waveform are known at 3PN~\cite{Bohe:2015ana}\footnote{And at 4PN for the dynamics only~\cite{Levi:2015ixa}.}, at the moment the effect is only consistently included at the leading 2PN order. An extension to higher orders is left for future work\footnote{We note that TaylorF2 models do include all known spin-square terms at 3PN (see Appendix~\ref{sec:taylorf2}). They are also present in the PhenomPNRT and SEOBNRT models. TEOBResumS was recently updated to include these next-to-leading terms~\cite{NagarMessinaRettegno2018}.}.

In the compact binary system we consider, the two material bodies, e.g. two neutron stars, with masses and spins $m_{A}$ and $\bs{S}_{A}$ for $A=1,2$, have spin-induced mass quadrupole moments $Q_{A,\mr{SS}}^{ij}=-\kappa_A S_A^{\langle i}S_A^{j\rangle}/m_A$ \cite{Poisson1998}, with $\kappa_{A}$ the quadrupole-monopole parameter. The case of two black holes is recovered for $\kappa_1=\kappa_2=1$, while for a neutron star $\kappa$ can be larger, of order 10 for hard EOSs.

The SEOB model should be modified as follows in order to take this into account at leading order. The SEOB effective Hamiltonian is related to the real Hamiltonian by (7.2) of~\cite{Barausse:2009xi}. Its structure is given by (5.70) of~\cite{Barausse:2009xi} as
\be
H_\mr{eff}=H_S+\beta^ip_i+\alpha\sqrt{\mu^2+\gamma^{ij}p_ip_j+\mc Q_4(p)}+H^\mr{BBH}_\mr{extra} \,,
\ee
where
\be
H^\mr{BBH}_\mr{extra}=\frac{1}{2r^3}(3n_in_j-\delta_{ij})\frac{\mu}{M}S_*^iS_*^j \,,
\ee
in terms of the total mass $M$, the reduced mass $\mu = M \eta$, and the spin combination
\be	
	S_*^i=\frac{m_2}{m_1} S_1^i+\frac{m_1}{m_2}S_2^i \,.
\ee
We refer to~\cite{Barausse:2009xi} for the notations and meaning of the other terms. We find that, for systems including neutron stars, the latter contribution should be replaced with
\begin{align}
	H^\mr{BNS}_\mr{extra} &= \frac{1}{2r^3}(3n_in_j-\delta_{ij}) \nnm\\
	& \; \cdot \bigg[\frac{\mu}{M}S_*^iS_*^j+(\kappa_1-1)\frac{m_2}{m_1}S_1^i S_1^j+(\kappa_2-1)\frac{m_1}{m_2}S_2^i S_2^j\bigg] \label{eq:HBNSextra}
\end{align}
where we chose to use the spin variables $\bs S_1$ and $\bs S_2$ instead of $\bs S_*$ and $\bs S=\bs S_1+\bs S_2$ to avoid the occurence of the mass difference $m_{1} - m_{2}$ in denominators. This expression generalizes the black hole case and reduces to it for $\kappa_{1} = \kappa_{2} = 1$.

In the waveform, at the leading 2PN order, the only contribution to consider will be in the mode $h_{22}$. The total leading-order spin-squared contribution, for aligned-spin circular orbits, is given by~\cite{Kidder:1995zr, Will:1996zj, Buonanno:2012rv}
\be
h_{22}^\tr{LO-SS}=-\frac{8\eta M^3\omega^2}{R}\sqrt{\frac{\pi}{5}}e^{-2i\Phi}\mc A \,,
\ee
with $\Phi$ the orbital phase, $\omega = \dot{\Phi}$ the orbital frequency and $R$ the distance to the observer. Here, for a BBH,
\be\label{eq:ABBH}
\mc A^\mr{BBH}=\frac{1}{M^2}\left(\frac{S_1^2}{m_1^2}+2\frac{S_1S_2}{m_1m_2}+\frac{S_2^2}{m_2^2}\right) \,,
\ee
while for a BNS, this should be replaced with
\be\label{eq:ABNS}
\mc A^\mr{BNS}=\frac{1}{M^2}\left(\kappa_1\frac{S_1^2}{m_1^2}+2\frac{S_1S_2}{m_1m_2}+\kappa_2\frac{S_2^2}{m_2^2}\right) \,.
\ee
The total spin-squared contribution to the flux also includes squares of spin-orbit terms, and reads~\cite{Kidder:1995zr, Will:1996zj, Poisson1998}
\be
\mc F^\tr{LO-SS}=\frac{32\eta^2}{5} v^{14} \left[2\mc A+\frac{1}{16M^2}\left(\frac{S_1}{m_1}-\frac{S_2}{m_2}\right)^2\,\right] \,,
\ee
with $v=(M\omega)^{1/3}$.

When translating the above for the EOB factorized waveform (see e.g. (17) in~\cite{Taracchini:2012ig}), spin contributions in the effective source, tail factors, and phase contribution would enter only at higher order, so that the only modification to consider is in the $\rho_{\ell m}$ factor for $\ell = 2$, $m = 2$. We have simply
\be\label{eq:rho22}
	\rho_{22}^{\textrm{LO-SS}} = \frac{1}{2}\mathcal{A} v^{4} \,,
\ee
with $\mathcal{A}$ the quantity given by~\eqref{eq:ABBH} for a BBH or~\eqref{eq:ABNS} for a BNS. 

\subsection{Waveform termination}

We now look for a suitable time $t_{\rm match}$ to stop the numerical integration and match with an effective post-merger waveform. In the point-mass model, \texttt{SEOBNRv4}, nonquasicircular corrections to the inspiral-plunge signal guarantee that the waveform peaks, and $t_{\rm match}$ is chosen to be the time of the peak amplitude $t_{\rm peak}^{\rm amp}$. Here, although we include nonquasicircular corrections, they are computed as in the BBH case and are not tuned to BNS simulations, so we have less control over the behavior of Eq.~(\ref{hlminspplunge}) in the late inspiral. We therefore choose the following definition for $t_{\rm match}$. Let $t_{\rm peak}^{\rm amp}$ be the earliest time when the amplitude $|h_{22}^{\textrm{insp-plunge}}|$ of the (2, 2) mode peaks, and let $t_{\rm peak}^{\rm freq}$ be the time when the frequency $\omega_{22}^{\rm insp-plunge}$ of the (2, 2) mode peaks. Then,
\begin{equation}
t_{\textrm{match}} = \min \left(t_{\textrm{peak}}^{\textrm{amp}},t_{\textrm{peak}}^{\textrm{freq}}\right)\,.\label{tmatch}
\end{equation}
For some waveforms, the amplitude does not peak. In this case, we choose $t_{\rm peak}^{\rm amp}$ to be the earliest time when the slope of the amplitude $\partial_t |h_{22}^{\rm insp-plunge}|$ reaches a minimum after having reached a peak.

In BNS simulations, the waveform rapidly decreases in amplitude in 1--2 gravitational-wave cycles after reaching peak amplitude. The BNS system then either undergoes prompt collapse or post-merger oscillations of the remnant. We do not attempt to model post-merger oscillations, and instead model an approximate peak emission and a subsequenct tapering to zero. This signal will be represented in terms of analytic functions for the amplitude $A^{\textrm{post-mrg}}(t)$ and phase $\phi^{\textrm{post-mrg}}(t)$ such that $h_{22}^{\textrm{post-mrg}}(t) = A^{\textrm{post-mrg}}(t) \exp{[i\phi^{\textrm{post-mrg}}(t)]}$. 

For the amplitude, we smoothly extend the waveform after $t_{\rm match}$ with a linear fit, and then taper the resulting amplitude. The linear extension is defined by
\begin{align}
\hat{A}(t)=\left\{\begin{array}{ll}
|h_{22}^{\rm insp-plunge}(t)|, & t \leq t_{\rm match}\,, \\
a + b (t - t_{\rm match}), & t > t_{\rm match}\,,
\end{array}\right.
\end{align}
where $a = |h_{22}^{\rm insp-plunge}(t_{\rm match})|$ and $b=\partial_t |h_{22}^{\rm insp-plunge}(t_{\rm match})|$. The tapering function is centered $15M$ after $t_{\rm match}$ and has a decay time $\tau = 2\pi / \omega_{\rm match}$ (where $\omega_{\rm match} = \omega_{22}^{\rm insp-plunge}(t_{\rm match})$) of one gravitational wave period. It is given by
\begin{equation}
W(t) = \frac{1}{1+\exp{[(t- t^{22}_{\textrm{match}}-15M)/\tau]}}\,.
\end{equation}
The final amplitude after windowing is then given by $|h_{22}(t)| = \hat A(t) W(t)$.

For the phase, we smoothly extend the waveform frequency such that it agrees with the inspiral frequency at a time $t_{\rm freq} = t_{\rm match} - 12M$ before the matching time $t_{\rm match}$, but then stretches out the frequency evolution such that it only approaches $\omega_{\rm match}$ asymptotically. We define this frequency evolution as
\begin{align}
\omega(t) = \omega_{\rm match} - \Delta\omega \exp\left[ -(t - t_{\rm freq}) / (12M) \right],
\end{align}
where $\Delta\omega = \omega_{\rm match} - \omega_{\rm freq}$. Integrating the frequency, and requiring continuity at $t_{\rm freq}$, results in the final expression for the phase
\begin{align}
\phi_{22}(t) = \left\{\begin{array}{ll}
\phi_{22}^{\textrm{insp-plunge}}(t), & t \leq t_{\rm freq} \,, \\
\phi_{\rm freq} + \omega_{\rm match}(t - t_{\rm freq}) + 12M \Delta\omega \left\{ \right. & \\
  \left. \exp\left[ -(t - t_{\rm freq}) / (12M) \right] - 1\right\} \,,& t > t_{\rm freq} \, .
\end{array}\right.
\end{align}

Although this effective post-merger model has not been fit to NR BNS simulations, it is in reasonable qualitative agreement with the post-merger behavior of the two equal-mass nonspinning NR BNS simulations that were analyzed in Ref.~\cite{Hinderer:2016eia}. A more sophisticated, NR-informed model of the post-merger emission will be part of future investigations.

\subsection{Reducing the number of matter parameters}

After rescaling with the total mass $M$, the aligned-spin EOB waveform with dynamical tides depends on 13 intrinsic parameters: the mass ratio $q=m_B/m_A\leq1$, the two dimensionless spin components along the orbital angular momentum $\chi_{A,B}$, the two spin-induced quadrupole-monopole parameters $\kappa_{A,B}$, the two adiabatic quadrupolar tidal polarizabilities $\Lambda_2^{A,B}$, the two adiabatic octupolar tidal polarizabilities $\Lambda_3^{A,B}$, and the four $\ell=2,3$ fundamental $f$-mode angular frequencies $\omega_{0\ell}^{A,B}$. To reduce the dimensionality of the intrinsic parameter space, we use nearly EOS-independent fits (universal relations) for these parameters in terms of $\Lambda_2$ as discussed below. This reduces the number of matter parameters from ten to two.

Eq.~(15) of Ref.~\cite{YagiYunes2017} provides a fit for $\kappa$ as a function of $\Lambda_2$. This relation has been fit for a sample of EOSs and for tidal parameters in the range $1 \leq \Lambda_2\leq10^4$. This formula, however, diverges at small values of $\Lambda_2$, so we replace it with a polynomial function that approaches the Kerr value of $\kappa = 1$ at $\Lambda_2 = 0$. We require this extension to be continuous at $\Lambda_2 = 1$. The exact relation we use is
\begin{equation}
\kappa = \left\{\begin{array}{ll}
1 + f \Lambda_2 + g \Lambda_2^2 + h \Lambda_2^3\, , & 0 \leq \Lambda_2 \le 1 \\
e^{a + b \xi + c \xi^2 + d \xi^3 + e \xi^4}\, , & \Lambda_2 > 1
\end{array}\right.,
\end{equation}
where $\xi = \ln\Lambda_2$. The coefficients $\{a, b, c, d, e\} = \{ 0.194, 0.09163, 0.04812, -4.283\times 10^{-3}, 1.245\times 10^{-4} \}$ are from Table I of Ref.~\cite{YagiYunes2017}, and $\{f, g, h\} = \{0.42769, -0.32434, 0.11074\}$.

Eq.~(60) of Ref.~\cite{Yagi:2013sva} provides a fit for $\Lambda_3$ as a function of $\Lambda_2$ with an accuracy of about 10\% depending on the EOS. This relation has been fit for a sample of EOSs and for tidal parameters in the range $1 \leq \Lambda_2\leq10^6$. This formula, however, diverges at small values of $\Lambda_2$, so we replace it with a polynomial function that vanishes at $\Lambda_2 = 0$. We require this extension to be continuous at $\Lambda_2 = 10^{-2}$, and to fit the universal relation in the range $10^{-5}\leq\Lambda_2\leq10^{-2}$. The exact relation we use is
\begin{equation}
\Lambda_3 = \left\{\begin{array}{ll}
\Lambda_2 (f + g \Lambda_2 + h \Lambda_2^2)\, , & 0 \leq \Lambda_2 \leq 10^{-2} \\
e^{a + b \xi + c \xi^2 + d \xi^3 + e \xi^4}\, , & \Lambda_2>10^{-2}
\end{array}\right.,
\end{equation}
where $\xi = \ln\Lambda_2$, $\{a, b, c, d, e\} = \{-1.15, 1.18, 2.51\times 10^{-2}, -1.31\times 10^{-3}, 2.52\times 10^{-5}\}$ from Ref.~\cite{Yagi:2013sva}, and $\{f, g, h\}=\{0.440649, -34.632322, 1762.112913\}$.

Eq. (3.5) of Ref.~\cite{Chan:2014kua} gives relations for $\omega_{02}$ as a function of $\Lambda_2$ and $\omega_{03}$ as a function of $\Lambda_3$ to within a few percent error. For $\omega_{02}$, the fitting range used by Ref.~\cite{Chan:2014kua} was $0\leq \xi \leq 9$, and outside this range we require continuity. The relation we use is then
\begin{equation}
\omega_{02} = \left\{\begin{array}{ll}
f\, , & \xi < 0 \\
a + b\xi + c\xi^2 + d\xi^3 + e\xi^4\, , & 1 \le \xi \le 9 \\
g\, , & \xi > 9
\end{array}\right.,
\end{equation}
where $\{a, b, c, d, e\} = \{0.182, -6.836\times10^{-3}, -4.196\times10^{-3}, 5.215\times10^{-4}, -1.857\times10^{-5}\}$ from Ref.~\cite{Chan:2014kua}, and $\{f, g\}=\{0.182, 0.161\}$. For $\omega_{03}$, the fit is given in terms of $\Upsilon=\ln\Lambda_3$ in the range $-1\leq \Upsilon \leq 10$, and outside this range we require continuity. The relation we use is then
\begin{equation}
\omega_{03} = \left\{\begin{array}{ll}
f\, , & \Upsilon < -1 \\
a + b\Upsilon + c\Upsilon^2 + d\Upsilon^3 + e\Upsilon^4\, , & -1 \le \Upsilon \le 10 \\
g\, , & \Upsilon > 10
\end{array}\right.,
\end{equation}
where $\{a, b, c, d, e\} = \{0.2245, -1.5\times10^{-2}, -1.412\times10^{-3}, 1.832\times10^{-4}, -5.561\times10^{-6}\}$ and $\{f, g\}=\{0.2379, 0.1165\}$.

With these relations the waveform only depends on the five intrinsic parameters $\bx=\{q, \chi_A, \chi_B, \Lambda_2^A,\Lambda_2^B\}$. 
We note that this list does not include the total mass $M$ of the system. The point-mass part of the model is scale-invariant, and the tidal corrections only depend on $X_A=1/(1+q)$, $X_B=q/(1+q)$ and $\Lambda_2^{A,B}$. Finally, the simple model of the post-merger signal that we employ rescales with $M$ as well. It is not clear at what point in the transition from the inspiral to the post-merger this approximation breaks down. For example, whether the merging binary undergoes prompt collapse or forms a hypermassive remnant depends sensitively on the total mass. NR simulations will be needed to determine when after the merger this approximation is no longer valid.

\section{Surrogate model}
\label{sec:surrogate}

In this section we describe how we decompose the EOB waveform into smooth, slowly-varying functions and train a surrogate model for these functions. We will work with the frequency-domain waveform $\tilde h(Mf; \bx)$. This is favorable for data analysis which is usually done in the frequency domain. It also allows us to extend the model down to arbitrarily low frequencies with an analytic, frequency-domain model. Unfortunately, Fourier transforming a finite length waveform leads to a surrogate with more noise. We will show below, however, that sufficient filtering can solve this problem. The \texttt{SEOBNRv4T} waveform model, as implemented, works for mass ratios in the range $q \in [1/3, 1]$, spins $\chi_{1,2} \in [-0.5, 0.5]$ and tidal parameters $\Lambda_{1,2} \in [0, 5000]$. Our surrogate of this model, \texttt{SEOBNRv4T\_surrogate}, will be valid for the same range of parameters.

\subsection{Decomposition of the waveform}

\begin{figure}[htb]
\centering
\includegraphics[width=0.49\textwidth]{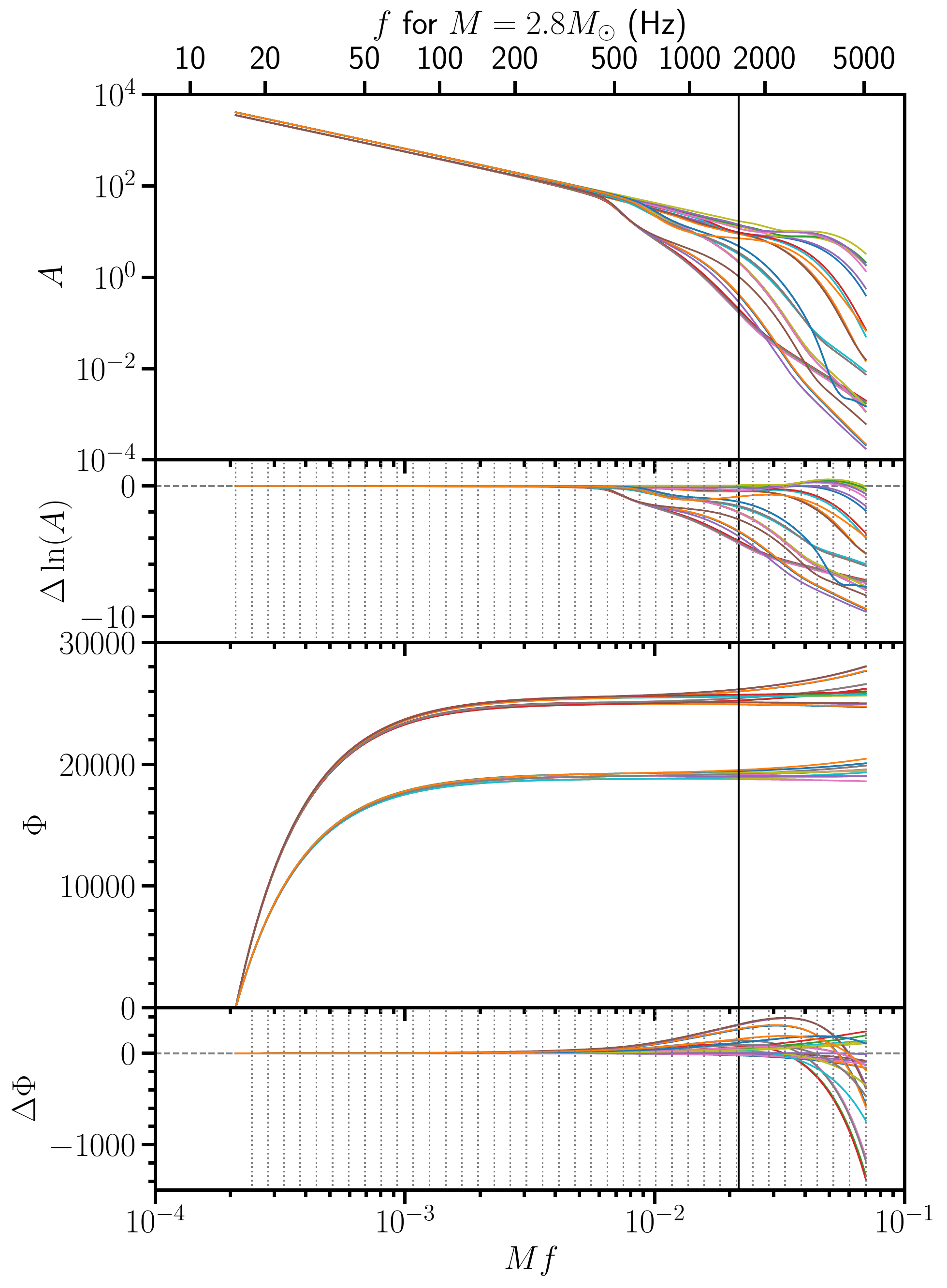}
\caption{EOB waveforms $\tilde h(Mf; \bx)$ for the 32 corners of parameter space. The waveforms are filtered and Fourier transformed as described in Sec.~\ref{sec:condition}. Top two panels: the amplitude $A(Mf; \bx)$ and residual $\Delta\ln A(Mf; \bx)$ relative to TaylorF2 as defined in Eq.~\eqref{eq:ampres}. Bottom two panels: the phase $\Phi(Mf; \bx)$ and residual $\Delta\Phi(Mf; \bx)$ relative to TaylorF2 as defined in Eq.~\eqref{eq:phaseres}. Vertical dashed lines represent the frequency nodes $MF_j$ where the residuals are interpolated as functions of $\bx$ using Gaussian process regression. The vertical solid line at $Mf_{\rm ISCO} = 1/(6^{3/2}\pi) \approx 0.022$ is the gravitational-wave frequency at the Schwarzschild ISCO, and the top axis is the frequency for a binary with a total mass of $2.8M_\odot$.}
\label{fig:hoff}
\end{figure}

Because the waveform $\tilde h(Mf; \bx)$ is an oscillatory function of $Mf$ and $\bx$, the waveform is usually decomposed into an amplitude $A(Mf; \bx)$ and phase $\Phi(Mf; \bx)$ as $\tilde h(Mf; \bx) = A(Mf; \bx) e^{i\Phi(Mf; \bx)}$. The amplitude and phase are smoother, mostly monotonic functions of frequency. 
This can be seen in Fig.~\ref{fig:hoff} where we show the waveforms for the 32 corners of the 5-dimensional parameter space. Unfortunately, the amplitude and phase still span a wide range of values. The phase, for example, spans about $10^4$~rad between waveforms with different parameters $\bx$ (see Fig.~\ref{fig:hoff}). To avoid systematic errors in the tidal parameters, for example, we need phase errors of $\lesssim 1$~rad over most of this frequency range, leading to a requirement on the fractional interpolation error of $\lesssim 10^{-4}$~rad. This is a difficult requirement to achieve for 5-dimensional interpolation. 

For aligned-spin waveforms, we can solve this problem by using the fact that the waveform can be approximated with the analytic TaylorF2 waveform $\tilde h_{\rm F2}(Mf; \bx) = A_{\rm F2}(Mf; \bx) e^{i\Phi_{\rm F2} (Mf; \bx)}$. This allows us to write the EOB amplitude and phase in terms of small residuals, $\Delta\ln(A)(Mf; \bx)$ and $\Delta\Phi(Mf; \bx)$, relative to TaylorF2
\begin{align}
\Delta\ln(A)(Mf; \bx) &= \ln\left(\frac{ A(Mf; \bx) }{ A_{\rm F2}(Mf; \bx)}\right), \label{eq:ampres}\\
\Delta\Phi(Mf; \bx) & = \Phi(Mf; \bx) - \Phi_{\rm F2}(Mf; \bx), \label{eq:phaseres}
\end{align}
such that
\begin{align}
\tilde h(Mf; \bx) = \tilde h_{\rm F2}(Mf; \bx) e^{\Delta\ln(A)(Mf; \bx) + i  \Delta\Phi(Mf; \bx)}.
\label{eq:hdecomp}
\end{align}
The exact functional form of the TaylorF2 waveform we use is given in Appendix~\ref{sec:taylorf2}, and the residuals are shown in Fig.~\ref{fig:hoff}. 

For the amplitude residual, we use a log-ratio instead of a ratio because it guarantees that interpolation errors will not lead to a negative amplitude for the reconstructed waveform (Eq.~\eqref{eq:hdecomp}). In addition, because the waveform amplitude spans several orders of magnitude at high frequencies, the log-ratio better captures this behavior. Comparing the phase $\Phi$ and phase residual $\Delta\Phi$ in Fig.~\ref{fig:hoff}, we find that the range in $\Delta\Phi$ is a few orders of magnitude smaller than the range in $\Phi$ except at very high frequencies. 

Finally, we have found that the functions $\Delta\ln A(Mf; \bx)$ and $\Delta\Phi(Mf; \bx)$ sometimes vary rapidly as a function of the tidal parameters $\Lambda_1$ and $\Lambda_2$ for tidal parameters in the range [0, 1000]. We therefore perform a change of variables that stretches out the parameter space for small values of the tidal parameters:
\begin{equation}
\xi_{A, B} = \log_{10}\left(\frac{\Lambda_{A, B}}{100} + 1\right).
\end{equation}
With $\bx=\{q, \chi_A, \chi_B, \xi_A,\xi_B\}$, the functions $\Delta\ln A(Mf; \bx)$ and $\Delta\Phi(Mf; \bx)$ are smoother functions of the parameters, making them easier to fit.

\subsection{Conditioning the training set waveforms}
\label{sec:condition}

The accuracy of the final frequency-domain surrogate depends on how well the finite-length, numerical waveform is Fourier transformed and filtered to remove numerical artifacts. We now describe the procedure to condition the training-set waveforms used to construct the surrogate.

We evaluate the EOB waveform with a starting frequency of $Mf_{{\rm win}, i}=0.000197$, equivalent to a physical frequency of 20~Hz for a binary with total mass $M=2M_\odot$. In order to take a discrete Fourier transform, we window the start of the waveform with a Planck window~\cite{McKechanRobinsonSathyaprakash2010} in the interval $[Mf_{{\rm win}, i}, Mf_{{\rm win}, f}] = [0.000197, 0.00021]$ to reduce Gibbs oscillations. The end of the waveform has zero amplitude, so the end does not need to be windowed. We then resample the waveform with a spacing $\Delta t/M = 5$ and pad the end of the waveform with zeros such that all waveforms in the training set have the exact same time samples. After evaluating the discrete Fourier transform, we calculate the residuals $\Delta\ln(A)(Mf)$ and $\Delta\Phi(Mf)$ between the EOB and TaylorF2 waveforms using Eqs.~\eqref{eq:ampres} and~\eqref{eq:phaseres}. 

Waveforms have free time and phase parameters $t_c$ and $\phi_c$, and in the frequency-domain this means that one can freely add a linear term $\phi_c + 2\pi (Mf) (t_c/M)$ to the phase $\Phi(Mf)$. We use this freedom to match the EOB waveform to the analytic TaylorF2 waveform near the starting frequency. We do this by subtracting a linear fit to $\Delta\Phi(Mf)$ at the beginning of the waveform in the window $[Mf_{{\rm fit}, i}, Mf_{{\rm fit}, f}] = 0.00021[1, 1.05]$. At $Mf=Mf_{{\rm fit}, i}$, the resulting phase residual $\Delta\Phi(Mf)$ is zero and has zero slope, guaranteeing that the surrogate smoothly matches to TaylorF2 below this frequency.

The resulting waveforms still have some remaining Gibbs oscillations which can be seen as small-amplitude, high-frequency oscillations in the residuals $\Delta\ln(A)(Mf)$ and $\Delta\Phi(Mf)$. These come from two sources. The first is the fact that the Planck window at the beginning of the waveform was not sufficiently long. We could make this window longer, but that would require us to start the surrogate at a higher frequency. The second source comes from the end of the waveform where the amplitude rapidly drops to zero amplitude during the $\sim 1$ cycle after the peak amplitude. Some contribution to the Gibbs oscillations at high frequencies is therefore a genuine feature of the EOB model. We reduce these oscillations in the frequency domain using a moving average filter centered on $Mf$ with interval $[Mf(1-\alpha), Mf(1+\alpha)]$. We use a width of $\alpha=0.1$ for the amplitude residual and $\alpha=0.05$ for the phase residual. Smoothing these oscillations makes it significantly easier to interpolate the amplitude and phase residuals as functions of $Mf$ and $\bx$. 

Finally, we truncate the residuals outside the interval $[Mf_{{\rm trunc}, i}, Mf_{{\rm trunc}, f}] = [0.00021, 0.07]$. 
We note that the gravitational-wave frequency at the Schwarzschild innermost stable circular orbit (ISCO) is $Mf_{\rm ISCO} = 1/(6^{3/2}\pi) \approx 0.022$, and higher frequencies represent only the last $\sim 1$ cycle of the waveform during the merger. Thus, the high-frequency cutoff of $Mf_{{\rm trunc}, f}=0.07$ is a very conservative upper bound for data-analysis purposes. However, truncating the waveform at lower frequencies can visibly effect the shape of some waveforms if inverse Fourier transformed back into the time domain.

\begin{figure}[htb]
\centering
\includegraphics[width=0.49\textwidth]{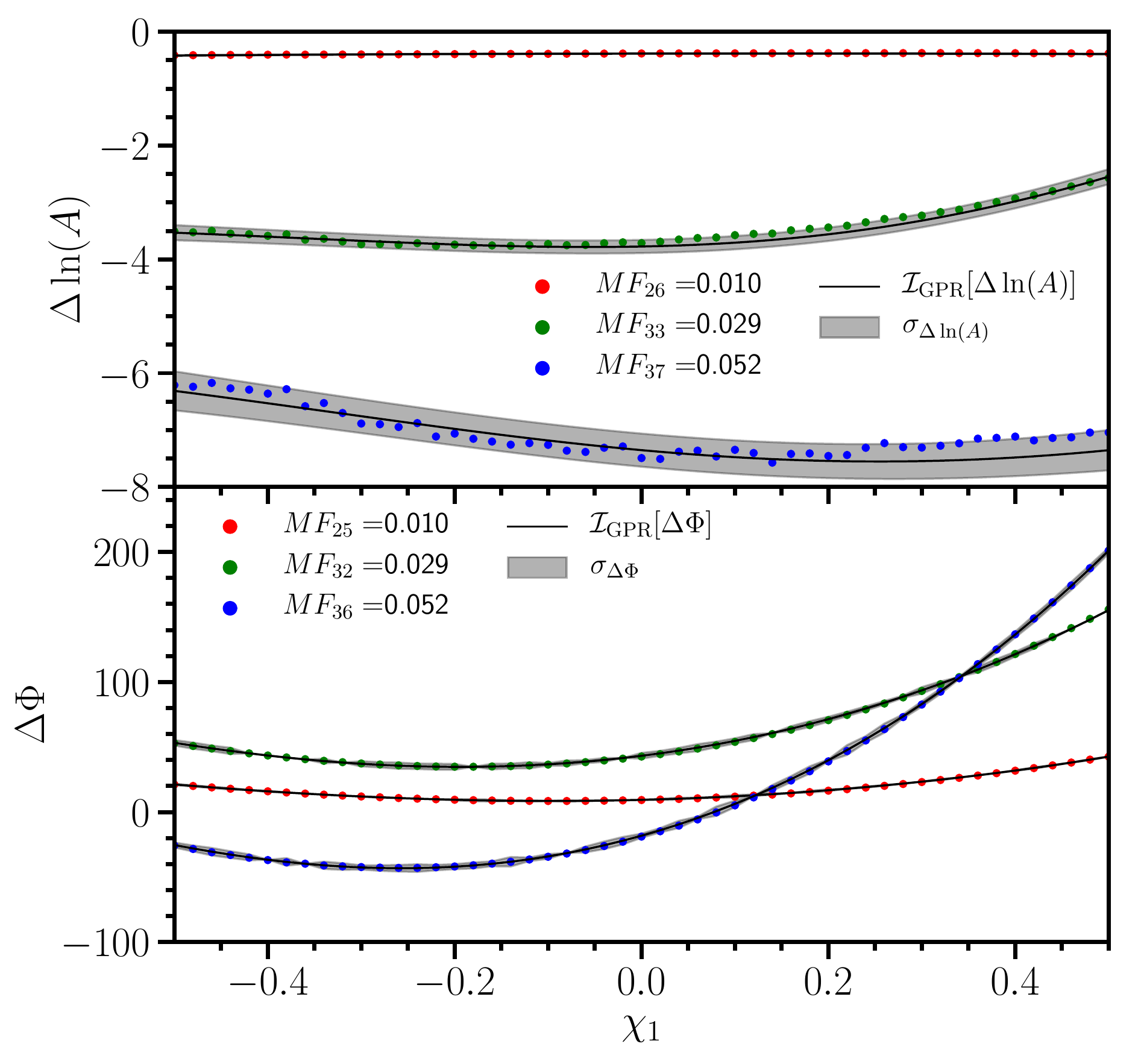}
\caption{Cross sections of the amplitude and phase residuals as functions of $\chi_1$ at three nodes $MF_j$. Test-set waveforms are generated for 51 values of $\chi_1$, shown as dots, with the other parameters held constant at $\{q, \chi_2, \Lambda_1, \Lambda_2\} = \{0.6, 0.2, 2000, 1000\}$. Also shown are the interpolated values for the final surrogate, discussed in Sec.~\ref{sec:accuracy}, using Gaussian process regression (solid black) and the corresponding $1\sigma$ uncertainty estimates (shaded gray region).}
\label{fig:dhofs}
\end{figure}

To validate that the waveforms are sufficiently conditioned, we plot cross-sections of the residuals in Fig.~\ref{fig:dhofs} for fixed frequencies as functions of one of the waveform parameters $\chi_1$. The fact that the residuals are smooth functions of the waveform parameters indicates that most numerical noise has been removed. The exception is the noisy amplitude residual at high frequencies, $\Delta\ln(A)(Mf=0.052; \bx)$, where the amplitude is very small and dominated by numerical noise. 

\subsection{Spline interpolation for frequency $f$}

We now seek to interpolate the conditioned residuals $\Delta\ln(A)(Mf; \bx)$ and $\Delta\Phi(Mf; \bx)$. We begin by choosing a method for interpolating as a function of $Mf$ for fixed $\bx$. The majority of previous papers on gravitational-wave surrogates have used an orthonormal basis of global functions $\hat e_i(Mf)$ for interpolating as a function of frequency (or time)~\cite{Puerrer2014, Puerrer2015, FieldGalleyHesthaven2014, LackeyBernuzziGalley2017, BlackmanFieldGalley2015, BlackmanFieldScheel2017a, BlackmanFieldScheel2017b}. This type of surrogate, built from a reduced basis, is usually referred to as a reduced order model. For a generic function $g(Mf; \bx)$, this decomposition can be written as $g(Mf; \bx) \approx \sum_{i=1}^N c_i(\bx) \hat e_i(Mf)$. The coefficients $c_i(\bx)$ are then interpolated as functions of $\bx$~\cite{Puerrer2014, Puerrer2015}. Alternatively, using the empirical interpolation method~\cite{Barrault2004, Chaturantabut2010, FieldGalleyHesthaven2014}, one can re-express these basis functions $\hat e_i(Mf)$ in terms of empirical interpolating functions $B_j(Mf)$ and the value of $g(Mf; \bx)$ at empirical nodes $MF_j$ as $g(Mf; \bx) \approx \sum_{j=1}^N B_j(Mf) g(MF_j; \bx)$. The location of these nodes $MF_j$ is then optimized to minimize interpolation errors.

For the problem here, we have found that global basis functions are not necessarily the optimal solution. The residuals are small and smooth at low frequencies and large and noisy at high frequencies. If using the empirical interpolation method with global basis functions, for example, errors in evaluating $g(MF_j;\bx)$ at high frequencies can propagate to large errors between the nodes $MF_j$ at low frequencies. Instead, we find that spline interpolation works significantly better. We use 40 frequency nodes $MF_j$ log-spaced in the interval $[Mf_{{\rm trunc}, i}, Mf_{{\rm trunc}, f}]$ (see Fig.~\ref{fig:hoff}). We evaluate the residuals $\Delta\ln(A)(MF_j; \bx)$ and $\Delta\Phi(MF_j; \bx)$ at these nodes as discussed below. We then interpolate between these frequencies using third-order splines. The local, third-order polynomials, that are only connected by the requirement of smoothness, do not propagate high-frequency errors down to low-frequency errors as significantly as do global basis functions.

\subsection{Gaussian process regression for parameters $\bx$}

Next we choose a method to interpolate the residuals $\Delta\ln(A)(MF_j; \bx)$ and $\Delta\Phi(MF_j; \bx)$ at each of the frequency nodes $MF_j$ as a function of the five waveform parameters $\bx$. Most multivariate interpolation techniques (e.g. tensor spline or Chebyshev interpolation) require a function to be sampled on a rectangular grid, and thus suffer from the curse of dimensionality: the number of samples grows exponentially with the dimension $d$ ($N^d$ samples for $N$ points per dimension). For our 5-dimensional problem, $10^5$ waveform evaluations are needed for only 10 samples per parameter. At a starting frequency of 20~Hz, over an hour is needed to evaluate a single \texttt{SEOBNRv4T} waveform on a standard CPU, so this is at the limit of what is reasonable. However, there are other methods that do not require a rectangular grid, and this allows us to choose more efficient experimental designs as discussed in Sec.~\ref{sec:design} below. The method we choose is GPR~\cite{RasmussenWilliams2006}.

In GPR, the values of a function $g(\bx)$ at the points $\bx$ are assumed to be a realization of a Gaussian process
\begin{equation}
g(\bx) \sim \mathcal{GP}(m(\bx), k(\bx, \bx')),
\end{equation}
where $m(\bx)$ is a mean function at each point $\bx$ and $k(\bx, \bx')$ is a covariance function between the points $\bx$ and $\bx'$.
When modeling data with GPR, one often attempts to subtract out the mean by first fitting the data with a parameterized function. Then, the residual is represented by a zero-mean Gaussian process. The covariance $k(\bx, \bx')$ is described by a kernel function with tunable hyperparameters. For the problem here, we have already subtracted the TaylorF2 waveform from the EOB waveform, so we model the residual in terms of a zero-mean Gaussian process. 

Explicitly, we are interested in predicting the function value $y_*$ at the point $\bx_*$ given the sampled function values $y_i$ at the points $\bx_i$.  In a zero-mean Gaussian process, this is represented by the following multivariate Gaussian distribution
\begin{equation}
\label{eq:gaussian}
\begin{bmatrix}
y_i \\
y_* \\
\end{bmatrix}
\sim \mathcal{N}
\left({\bm 0}, 
\begin{bmatrix}
K & K_*^T \\
K_* & K_{**} \\
\end{bmatrix}
\right),
\end{equation}
where $K_{ij} = k(\bx_i, \bx_j)$ is a matrix, $K_{*i} = k(\bx_i, \bx_*)$ is a vector, and $K_{**} = k(\bx_*, \bx_*)$ is a scalar.

From Eq.~\eqref{eq:gaussian}, the conditional probability for $y_*$ given the training set examples $y_i$ and kernel hyperparameters $\btheta$ is also a Gaussian
\begin{equation}
p(y_* | \bx_i, \bx_*, y_i, \btheta) = \mathcal{N}(\bar y_*, {\rm var}(y_*)),
\end{equation}
where the mean and variance are
\begin{align}
\label{eq:mean}
\bar y_* &= K_{*i} (K^{-1})_{ij} y_j, \\
\label{eq:var}
{\rm var}(y_*) &= K_{**} - K_{*i} (K^{-1})_{ij} K_{*j}.
\end{align}
Eq.~\eqref{eq:mean} is the estimate of the function, and Eq.~\eqref{eq:var} is the estimate of the uncertainty.

We use radial kernels, which express the covariance in terms of a distance $r$ between points
\begin{equation}
r^2 = (\bx - \bx')^T M (\bx - \bx'),
\end{equation}
where we choose the matrix $M$ to be diagonal
\begin{equation}
M = {\rm diag}(\ell_1^{-2}, \ell_2^{-2}, \dots, \ell_d^{-2}).
\end{equation}
The tunable hyperparameters ${\bf \ell}$ represent the length scale over which the function $g(\bx)$ varies in each coordinate. 

We try two classes of radial kernels. The first is the Mat\'{e}rn kernel
\begin{equation}
k_{\rm radial}(r) = \frac{2^{1-\nu}}{\Gamma(\nu)} \left(\sqrt{2\nu} r\right)^\nu K_\nu \left(\sqrt{2\nu} r \right),
\end{equation}
where $K_\nu(x)$ is a modified Bessel function. The value of $\nu$ parameterizes the smoothness of the Gaussian process, and is $k$ times mean-square differentiable if $\nu>k$~\cite{RasmussenWilliams2006}. For half-integer values of $\nu$, this kernel has a computationally cheap form without special functions, and we have had good results with $\nu=5/2$, resulting in a twice-differentiable function. The $\nu=5/2$ kernel is
\begin{equation}
k_{\rm radial}(r) = \left(1+\sqrt{5}r + \frac{5r^2}{3}\right) \exp\left(-\sqrt{5}r\right).
\end{equation}
The second class we try is the more common squared exponential kernel that results in an infinitely differentiable function~\cite{RasmussenWilliams2006}
\begin{equation}
k_{\rm radial}(r) = e^{-r^2/2}.
\end{equation}

Our final kernel takes the form
\begin{equation}
k(\bx_i, \bx_j) = \sigma_f^2 k_{\rm radial}(r) + \sigma_n^2 \delta_{ij},
\end{equation}
where $\sigma_f$ is a scale factor that describes the range of values that $g(\bx)$ takes over the domain, and $\sigma_n$ is a noise parameter. The white noise kernel $\sigma_n^2 \delta_{ij}$ (also called a nugget) parameterizes the noise in the data $y_i$. In our case, the training set waveforms have numerical noise that we will estimate by optimizing the hyperparameters. The full set of hyperparameters is now  $\btheta = \{\sigma_f, \ell_q, \ell_{\chi_1}, \ell_{\chi_2}, \ell_{\xi_1}, \ell_{\xi_2}, \sigma_n\}$ for our parameter space $\bx = \{q, \chi_1, \chi_2, \xi_1, \xi_2\}$.

In order to estimate the hyperparameters, we use the above assumption (Eq.~\eqref{eq:gaussian}) that the joint distribution of 
the data $y_i$ is a multivariate Gaussian with the following distribution:
\begin{equation}
\ln p(y_i | \bx_i, {\bm \theta}) = -\frac{1}{2} y_i (K^{-1})_{ij} y_j - \frac{1}{2} \ln |K| - \frac{d}{2} \ln 2\pi.
\end{equation}
This is the log-likelihood for $y_i$ given the hyperparameters ${\bm \theta}$, and we can find the posterior for ${\bm \theta}$
given $y_i$ using Bayes' theorem
\begin{equation}
p({\bm \theta} | \bx_i, y_i) \propto p({\bm \theta}) p(y_i | \bx_i, {\bm \theta}).
\end{equation}
The prior $p({\bm \theta})$ is typically uniform and used to set the bounds on ${\bm \theta}$. One can sample this posterior if interested in the distribution of hyperparameters $\btheta$. However, for the problem here, we simply want the maximum posterior. We do this using the \texttt{gaussian\_process} module in the \texttt{scikit-learn} package~\cite{scikit-learn}. With these optimized hyperparameters, the final interpolating function is given by Eq.~\eqref{eq:mean} and its uncertainty by Eq.~\eqref{eq:var}.

\subsection{Surrogate waveform evaluation}

Given the GPR fits for the amplitude and phase residuals, we can now reconstruct the frequency-domain waveform. We label the set of GPR fits at the frequency nodes $MF_j$ by $\{\mathcal{I}_{\rm GPR}[\Delta\ln A_j](\bx)\}$ for the amplitude residuals and by $\{\mathcal{I}_{\rm GPR}[\Delta\Phi_j](\bx)\}$ for the phase residuals. The surrogates for the residuals $\Delta\ln A_S(Mf;\bx)$ and $\Delta\Phi_S(Mf;\bx)$ are constructed by interpolating between the nodes $MF_j$ with cubic splines:
\begin{align}
\Delta\ln A_S(Mf;\bx) &= \mathcal{I}_{\rm Spline}[\{\mathcal{I}_{\rm GPR}[\Delta\ln A_j](\bx)\}](Mf),\\
\Delta\Phi_S(Mf;\bx) &= \mathcal{I}_{\rm Spline}[\{\mathcal{I}_{\rm GPR}[\Delta\Phi_j](\bx)\}](Mf).
\end{align}
We set these functions to 0 below the first frequency node $MF_0 =  Mf_{{\rm trunc},i}$ so that the waveform transitions to TaylorF2 at lower frequencies. With the analytic expressions for $A_{\rm F2}(Mf;\bx)$ and $\Phi_{\rm F2}(Mf;\bx)$ and the interpolated expressions $\Delta\ln A_S(Mf;\bx)$ and $\Delta\Phi_S(Mf;\bx)$, the final surrogates for the amplitude and phase are
\begin{align}
\label{eq:amp_sur}
A_S(Mf;\bx) &= A_{\rm F2}(Mf;\bx) \exp\left[ \Delta\ln A_S(Mf;\bx) \right], \\
\label{eq:phase_sur}
\Phi_S(Mf;\bx) &= \Phi_{\rm F2}(Mf;\bx) + \Delta\Phi_S(Mf;\bx).
\end{align}
In physical units, for an inclination angle $\iota$, the $+$ and $\times$ polarizations of the waveform are (for $f>0$)
\begin{align}
\label{eq:hplus_sur}
\tilde h_+(f; \bx) &= \frac{1}{2}(1+\cos^2\iota) \frac{G^2 M^2}{c^5 d} A_S\left(\frac{GMf}{c^3}; \bx\right) \nonumber \\
& \times \exp\left[i \Phi_S\left(\frac{GMf}{c^3}; \bx\right)\right], \\
\label{eq:hcross_sur}
\tilde h_\times(f; \bx) &= \cos\iota \frac{G^2 M^2}{c^5 d} A_S\left(\frac{GMf}{c^3}; \bx\right) \nonumber \\
& \times \exp\left[i \Phi_S\left(\frac{GMf}{c^3}; \bx\right) + i \frac{\pi}{2}\right].
\end{align}

\subsection{Iterative construction of training set and surrogate}
\label{sec:design}

One of the main aims of this paper is to build a surrogate with as few waveform evaluations as possible. Using the freedom provided by GPR to sample waveforms at arbitrary locations, we try a set of designs more efficient than uniform grids. One such design is a Latin Hypercube Design (LHD)~\cite{McKayBeckmanConover1979}. An LHD with $N$ samples divides each of $d$ dimensions uniformly into $N$ grid points for a total of $N^d$ grid points. However, unlike a uniform grid, the $N$ values in each dimension are sampled exactly once instead of $N^{d-1}$ times. For an LHD there are $(N!)^d$ ways to choose these points, and we choose one randomly\footnote{There are additional ways to choose an LHD. A standard requirement is that the LHD be space-filling, meaning that the points are as far apart as possible from each other. (See Ref.~\cite{Husslage2011} for a review of methods for optimizing the placement of samples.) One such definition of space filling is that the chosen locations maximize the minimum Euclidean distance between any two samples. We have not experimented with these alternatives here.}. An LHD has the property that the samples are non-collapsing; a projection onto a subspace is still an LHD and no points are repeated in any dimension. This avoids wasting samples when one of the parameters has much less of an influence than the other parameters. 

\begin{figure}[htb]
\centering
\includegraphics[width=0.49\textwidth]{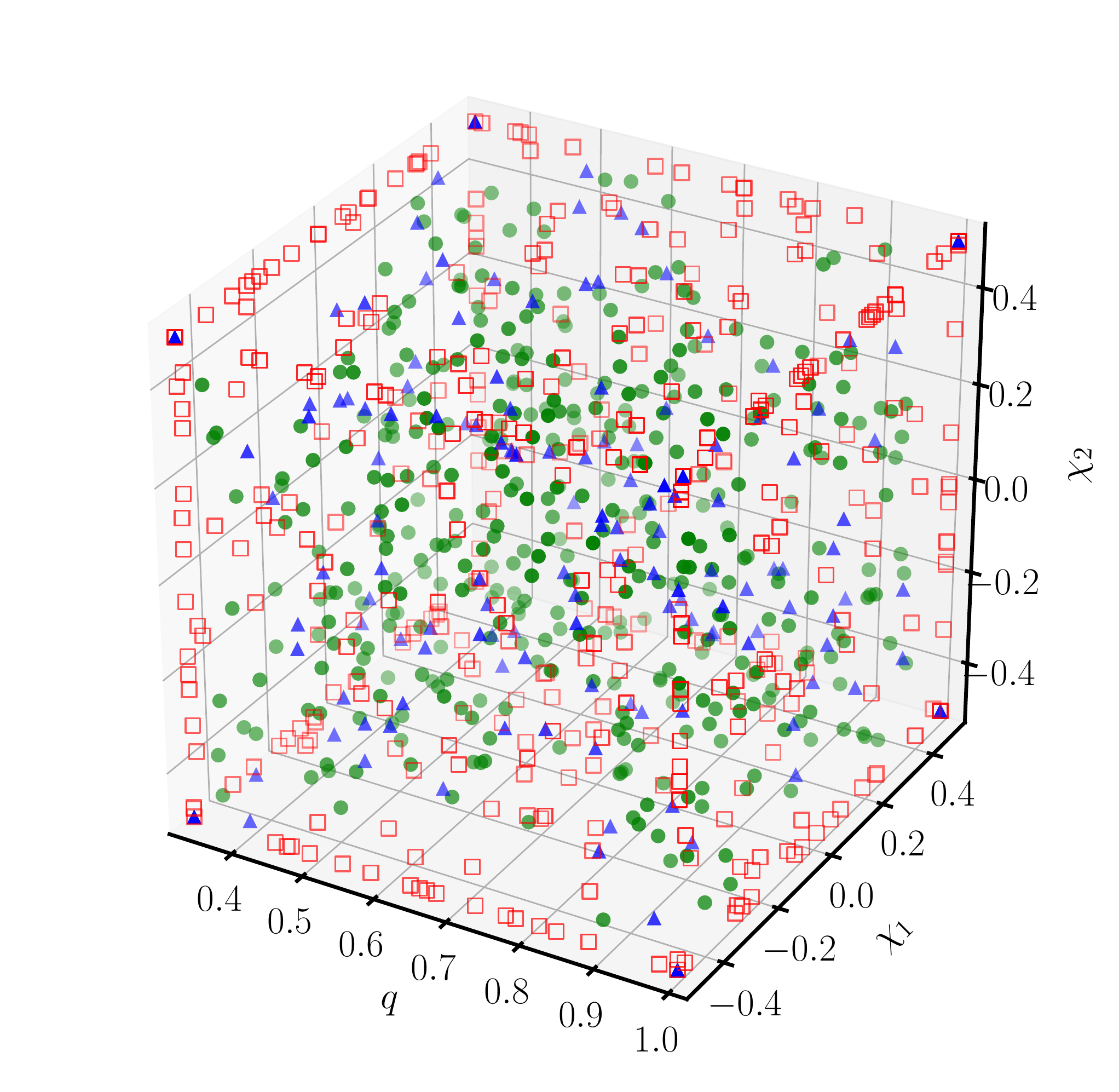}\\
\includegraphics[width=0.49\textwidth]{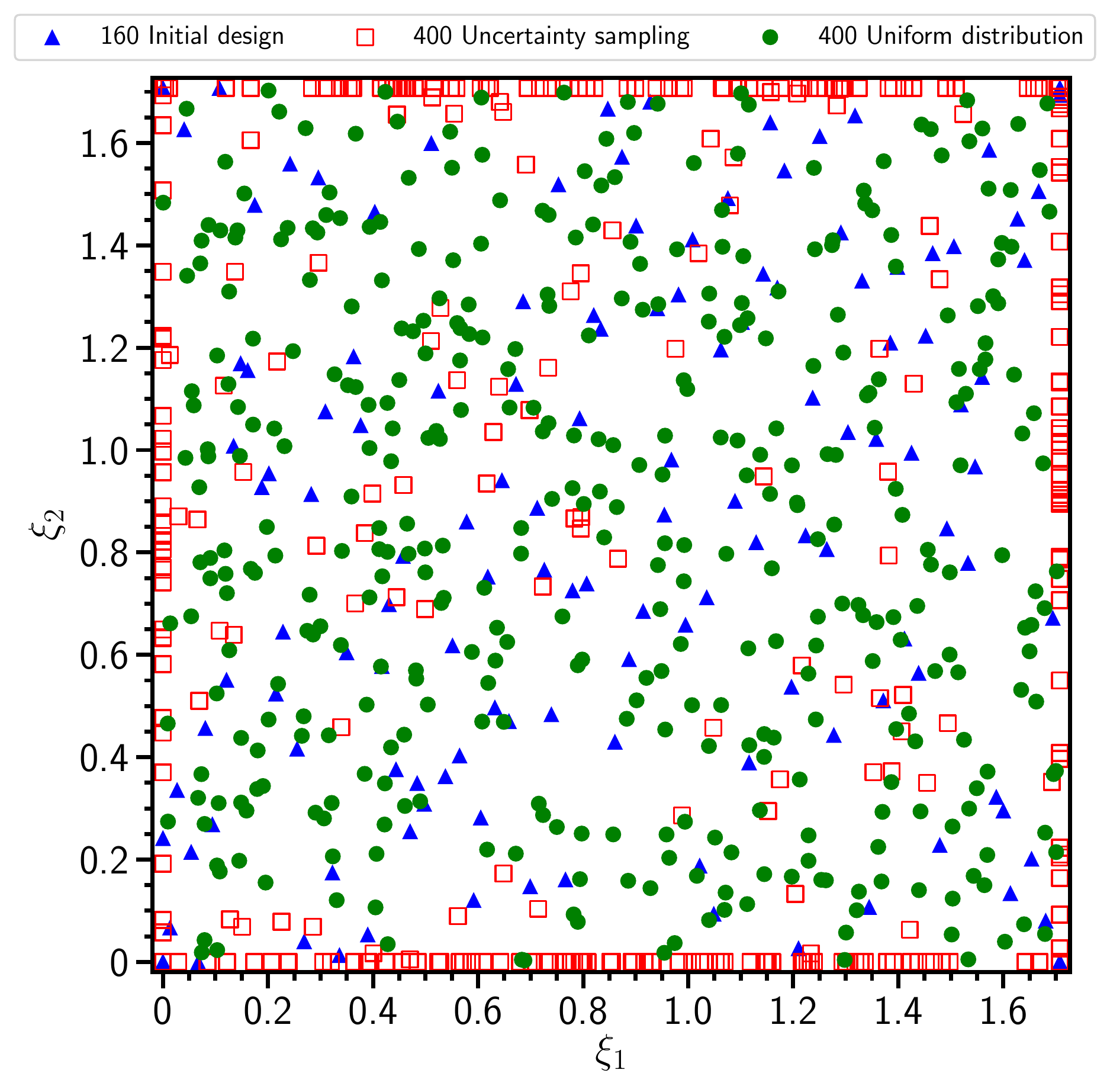}
\caption{Projection of the training-set parameters onto the three-dimensional subspace $\{q, \chi_1, \chi_2\}$ (top) and the two-dimensional subspace $\{\xi_1, \xi_2\}$ (bottom). The 160 blue triangles were used for the initial training set (32 corners and 128 LHD points). The 400 red squares were generated using uncertainty sampling from the GPR error estimate (Eq.~\eqref{eq:errorrms}) with the Mat\'{e}rn kernel. The 400 green circles were sampled uniformly from the parameter space.}
\label{fig:LHD}
\end{figure}

We build an initial training set with 128 waveforms sampled with an LHD for the five parameters $\bx = \{q, \chi_1, \chi_2, \xi_1, \xi_2\}$. 
In addition, we find empirically that the GPR uncertainty estimates (Eq.~\eqref{eq:var})
are largest at the corners of the parameter space, so we also sample the 32 corners. We construct our initial surrogate with these 160 waveforms shown in Fig.~\ref{fig:LHD}.

With this initial surrogate, we compare two methods for adding samples and improving the accuracy of the surrogate. The first is to simply sample the parameter space $\bx$ with a uniform distribution. We choose 400 samples shown in Fig.~\ref{fig:LHD}. The second method is sometimes referred to as uncertainty sampling~\cite{BrochuCoraDeFreitas2010}. In this method, we can choose new training-set samples by iteratively searching the parameter space for new points ${\bm x}$ that maximize some error criterion, then adding a new sample at that point. The quantity that we use is the root-mean-squared (RMS) phase error at the frequency nodes $MF_j$:
\begin{equation}
\epsilon_{\rm RMS}(\bx) = \sqrt{\frac{1}{N} \sum_{ \substack{j \\ MF_j \le 0.03} } [\sigma_{\Delta\Phi_j}(\bx)]^2},
\label{eq:errorrms}
\end{equation}
where $\sigma_{\Delta\Phi_j}(\bx)$ is the GPR estimate (Eq.~\eqref{eq:var}) of the uncertainty in $\Delta\Phi(MF_j, \bx)$ at node $MF_j$. We only include nodes below $Mf=0.03$ because there is very little signal at higher frequencies, and the RMS error would otherwise be dominated by higher frequencies. There are many alternative ways to construct a scalar that estimates overall waveform error. One alternative would be an approximation to the mismatch discussed in Sec.~\ref{sec:accuracy}. However, the quantity $\epsilon_{\rm RMS}(\bx)$ describes the most important quantity, phase, and is fast to compute. For a training set with $N$ samples, $\epsilon_{\rm RMS}(\bx)$ has $\sim N$ local maxima located in the voids between samples $\bx$. To find the global maximum, we use a basin hopping algorithm~\cite{WalesDoye1998, scipy:basinhopping} to avoid getting stuck in local maxima.

For low starting frequencies, EOB waveforms are expensive enough that we would like a method to efficiently choose $N_{\rm new}$ waveforms and evaluate them in parallel. We note that if we hold the hyperparameters $\btheta$ fixed, the GPR error estimate (Eq.~\eqref{eq:var}), and therefore $\epsilon_{\rm RMS}(\bx)$, only depends on the samples $\bx$ and not on the waveform data. The algorithm for choosing the $N_{\rm new}$ new points is as follows.

For $i = 1, \dots, N_{\rm new}$:
\begin{enumerate}

\item Construct the GPR error estimate $\epsilon_{\rm RMS}(\bx)$ (Eq.~\eqref{eq:errorrms}). In practice this can be done by specifying the samples $\bx$, the hyperparameters $\btheta$ from the initial surrogate, and dummy data for $\by$ since Eq.~\eqref{eq:errorrms} does not depend on $\by$.

\item Find the point $\bx_{\rm max}$ that maximizes $\epsilon_{\rm RMS}(\bx)$ over the parameter space. 

\item Add $\bx_{\rm max}$ to the list of samples: ${\bm x} \to [{\bm x}, {\bm x}_{\rm max}]$.

\end{enumerate}
These $N_{\rm new}$ waveforms can now be evaluated in parallel. An updated surrogate can then be constructed with re-optimized hyperparameters $\btheta$ using the $N + N_{\rm new}$ waveforms.

Fig.~\ref{fig:LHD} shows the parameters of the $N_{\rm new}=400$ new waveforms chosen by the uncertainty sampling method when using a Mat\'{e}rn kernel. We note that the edges of the parameter space are more often chosen than the inner region. We find this is true for both the Mat\'{e}rn and squared exponential kernels. Ref.~\cite{DoctorFarrHolz2017} found similar behavior for 1- and 2-dimensional problems. As discussed there, this results because the edges have fewer nearby samples than the interior, resulting in a larger error estimate. The uncertainty sampling algorithm compensates by adding samples near the edges where the error estimate is largest.

\begin{figure}[htb]
\centering
\includegraphics[width=0.49\textwidth]{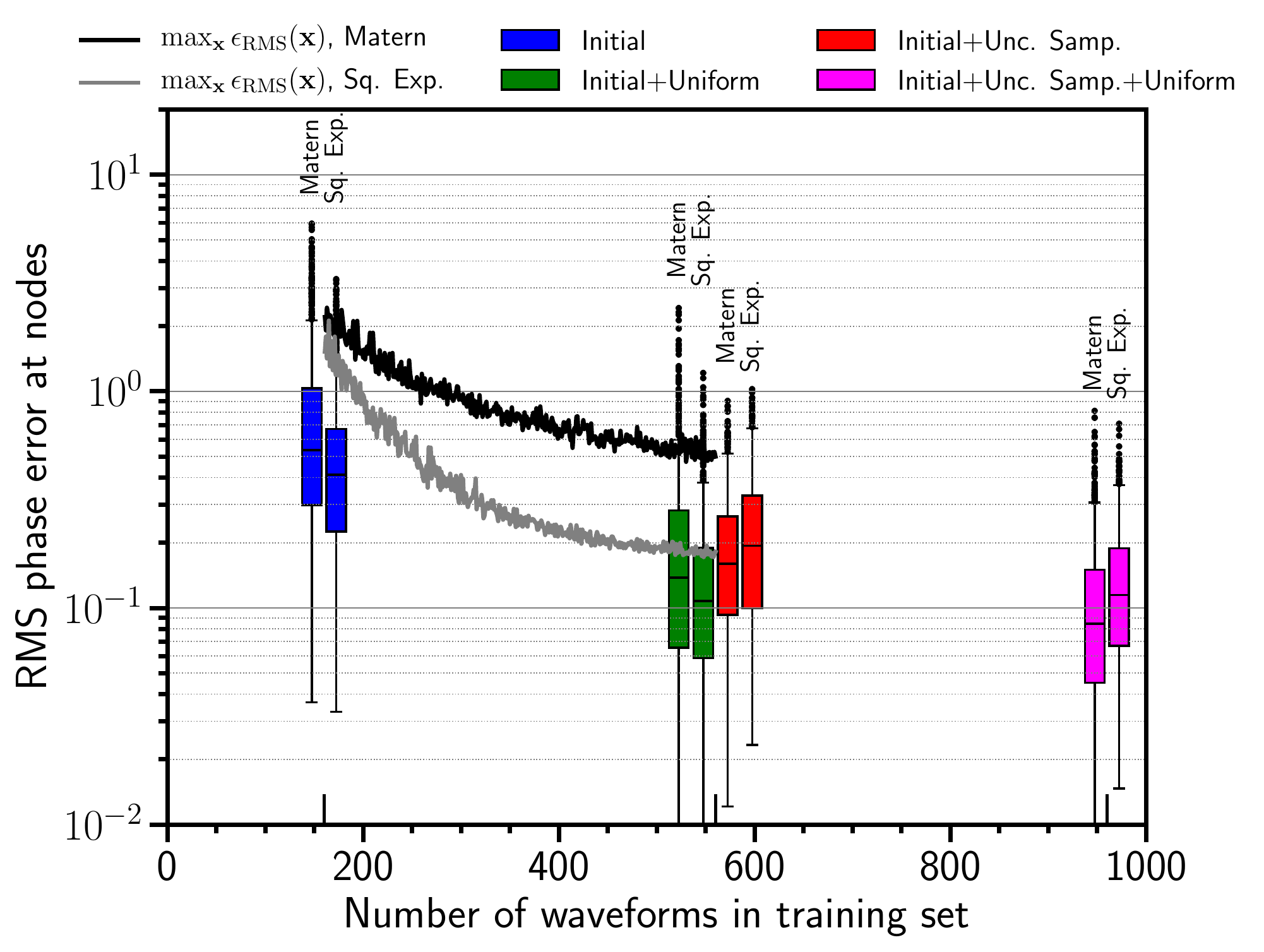}
\caption{Estimated $1\sigma$ RMS phase error $\epsilon_{\rm RMS}(\bx)$ (Eq.~\eqref{eq:errorrms}) at the frequency nodes $MF_j$ maximized over parameter space $\bx$ as a function of the number of samples. The black curve represents the estimated error using the Mat\'{e}rn kernel, and the gray curve represents the estimated error using the squared exponential kernel. 400 samples were added to the initial 160 samples for a total of 560 samples. Also shown are box plots of the maximum RMS phase errors at the nodes $MF_j \le 0.03$ for the surrogate compared to the test set of 1000 waveforms (Sec.~\ref{sec:accuracy}). The box plot represents the 25\%, 50\%, and 75\% quartiles. The whiskers contain all samples within 1.5 times the interquartile range. The black dots are outliers. Blue box plots: phase error from surrogates constructed using the initial design (160 corner and LHD waveforms) using either the Mat\'{e}rn or squared exponential kernel. Green box plots: surrogates constructed with the 160 initial waveforms and 400 uniformly sampled waveforms. Red box plots: surrogates constructed with the 160 initial waveforms and 400 waveforms chosen with uncertainty sampling. Magenta box plots: surrogates constructed with the 960 waveforms combining all sampling methods.}
\label{fig:uncsamp}
\end{figure}

Fig.~\ref{fig:uncsamp} shows the estimated RMS error $\epsilon_{\rm RMS}(\bx)$ maximized over the parameters $\bx$ for each new sample added to the training set. This is done for both the Mat\'{e}rn (black) and squared exponential (gray) kernels. For each kernel, we constructed an initial surrogate with optimized hyperparameters, then fixed the hyperparameters before choosing the 400 new samples using uncertainty sampling. The non-smoothness of the curves results because the basin-hopping algorithm does not always find the global maximum of $\epsilon_{\rm RMS}$ at each iteration. We find that increasing the number of waveforms from 160 to 560 decreases the global maximum of $\epsilon_{\rm RMS}$ by a factor of $\sim 4$ for the Mat\'{e}rn kernel and a factor of $\sim 10$ for the squared exponential kernel. In Section~\ref{sec:accuracy} we will examine how well this estimated error agrees with the true error.

Finally, after generating waveforms for the initial surrogate, the uncertainty sampling, and the uniform distribution, we combine all 960 waveforms. We then build the final surrogate with re-optimized hyperparameters.

\section{Performance of the surrogate}
\label{sec:performance}

\subsection{Accuracy}
\label{sec:accuracy}

The accuracy of the surrogate can be assessed by comparing it to a test set of waveforms. We generate 1000 waveforms with the original parameters sampled uniformly in the ranges $q \in [1/3, 1]$, $\chi_{1,2} \in [-0.5, 0.5]$ and $\Lambda_{1,2} \in [0, 5000]$. In Fig.~\ref{fig:uncsamp}, we show box plots of the RMS phase error of the test set waveforms for each of the 8 surrogate models. The first two surrogates are constructed using the initial design with either the Mat\'{e}rn or squared exponential kernel. The second two surrogates are constructed with the initial design and the 400 uniformly distributed samples and either the Mat\'{e}rn or squared exponential kernel. The next two surrogates use the initial design and the 400 samples chosen by the uncertainty sampling algorithm for each kernel. The final two surrogates use all 960 waveforms for each kernel. We find that for the Mat\'{e}rn kernel, the global maximum of $\epsilon_{\rm RMS}(\bx)$ is a reasonable estimate of the maximum test-set RMS phase errors. However, for the squared exponential kernel, the global maximum of $\epsilon_{\rm RMS}(\bx)$ significantly underestimates the true maximum test-set error. Additionally, we find that the surrogates constructed with the uniform distribution have smaller median test-set errors. However, the surrogates constructed with uncertainty sampling have smaller maximum test-set errors.

Another common measure of the surrogate model accuracy is the mismatch between the surrogate model and the test-set waveforms.
The mismatch represents the loss in signal-to-noise ratio that results from using the surrogate model $h_{\rm Sur}$ instead of the original waveform $h$. It is defined as the deviation from a perfect overlap after aligning the two waveforms using the time and phase free parameters $t_c$ and $\phi_c$:
\begin{equation}
\mathcal{M} = 1 - \max_{t_c, \phi_c} \frac{(h, h_{\rm Sur})} {\sqrt{(h, h) (h_{\rm Sur}, h_{\rm Sur})}}.
\end{equation}
The inner product between waveforms $h_1$ and $h_2$ is
\begin{equation}
(h_1, h_2) = 4 \Re \int_{f_{\rm low}}^{f_{\rm high}} \frac{\tilde h_1(f) \tilde h^*_2(f)} {S_n(f)} df,
\end{equation}
where the Fourier transformed waveforms are weighted by the noise power spectral density (PSD) $S_n(f)$ of the detector.

\begin{figure}[htb]
\centering
\includegraphics[width=0.49\textwidth]{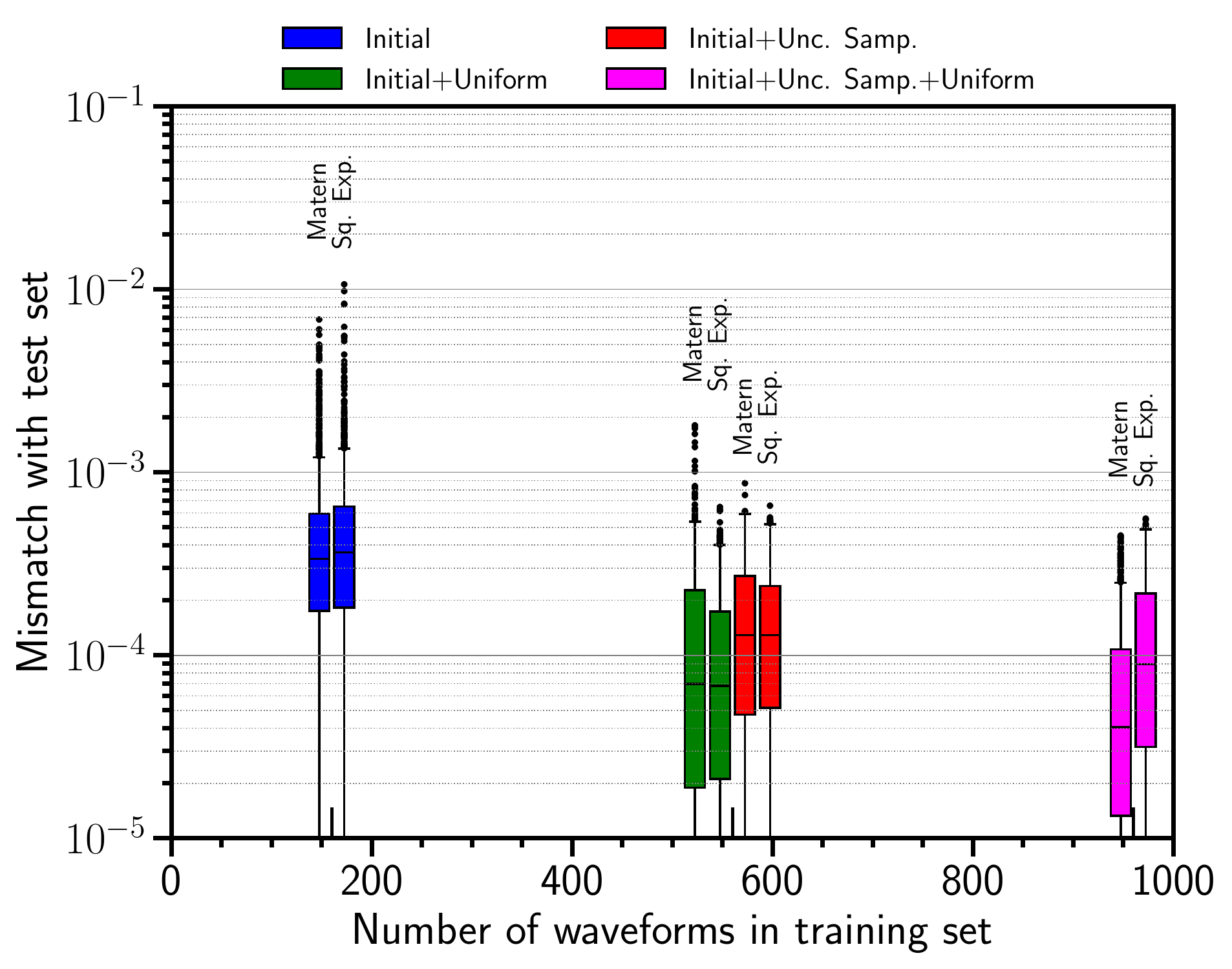}
\caption{
Mismatch between the surrogates and the 1000 test-set waveforms. The total mass for each waveform is $M=2.8M_\odot$, and the PSD is for the aLIGO design sensitivity configuration. Conventions for the box plots are the same as in Fig.~\ref{fig:uncsamp}.
}
\label{fig:mismatch}
\end{figure}

In Fig.~\ref{fig:mismatch}, we show box plots of the mismatches $\mathcal{M}$ between the surrogate and the 1000 test-set waveforms for each of our 8 surrogates. We use the design sensitivity aLIGO PSD~\cite{Aasi:2013wya} and a total mass of $M=2.8M_\odot$. Our integration bounds are $Mf_{\rm low} = 0.00021$ and $Mf_{\rm high} = 0.07$, corresponding to physical units of 15.2~Hz and 5076~Hz respectively. As we increase the number of samples in our training set, the mismatches decrease. However, it is less clear which kernel and sampling strategy is optimal. For our final surrogate, we simply choose the surrogate that has the smallest maximum and median mismatches, and this is the surrogate with the Mat\'{e}rn kernel and all 960 waveforms. This surrogate has a maximum mismatch of $4.5\times 10^{-4}$. 

\begin{figure}[htb]
\centering
\includegraphics[width=0.49\textwidth]{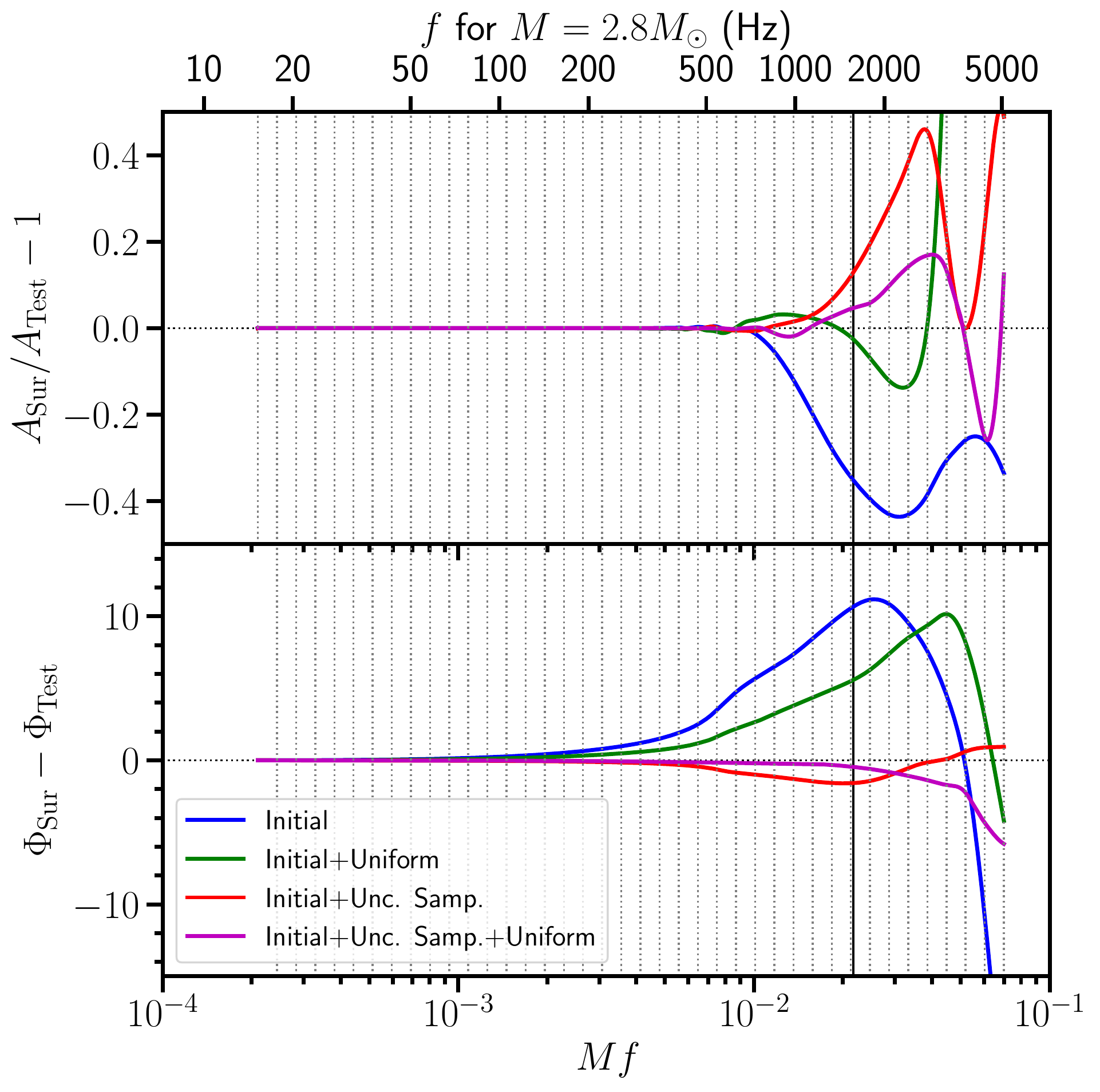}
\caption{Fractional amplitude and phase errors between the 4 surrogates with a Mat\'{e}rn kernel and the test-set waveform with the largest mismatch for each surrogate. The vertical solid line is the gravitational-wave frequency at the Schwarzschild ISCO, and the top axis is the physical frequency for a binary with a total mass of $2.8M_\odot$. Blue curve: surrogate with initial waveforms (160 corner and LHD waveforms). Green curve: surrogate with initial waveforms and 400 uniformly sampled waveforms. Red curve: surrogate with initial waveforms and 400 waveforms chosen with uncertainty sampling. Magenta curve: final surrogate with the 960 waveforms combining all sampling methods.}
\label{fig:maxmismatchfd}
\end{figure}

We also show the surrogate amplitude and phase errors as a function of frequency for the parameters with the largest mismatch relative to the test-set in Fig.~\ref{fig:maxmismatchfd}. As we add waveforms to the training set with the various sampling methods, the errors generally decrease. However, above a frequency of $Mf \sim 0.03$, the amplitude of the training set waveforms are noisy (see Figs.~\ref{fig:hoff} and~\ref{fig:dhofs}), and it is thus difficult to produce a good fit. This is not a problem, however, because aLIGO is insensitive to the waveform morphology at such high frequencies. With our final surrogate, we are able to achieve a maximum phase error of $\lesssim 1$~radian over the entire frequency range relevant to aLIGO.

\begin{figure*}[htb]
\centering
\includegraphics[width=0.99\textwidth]{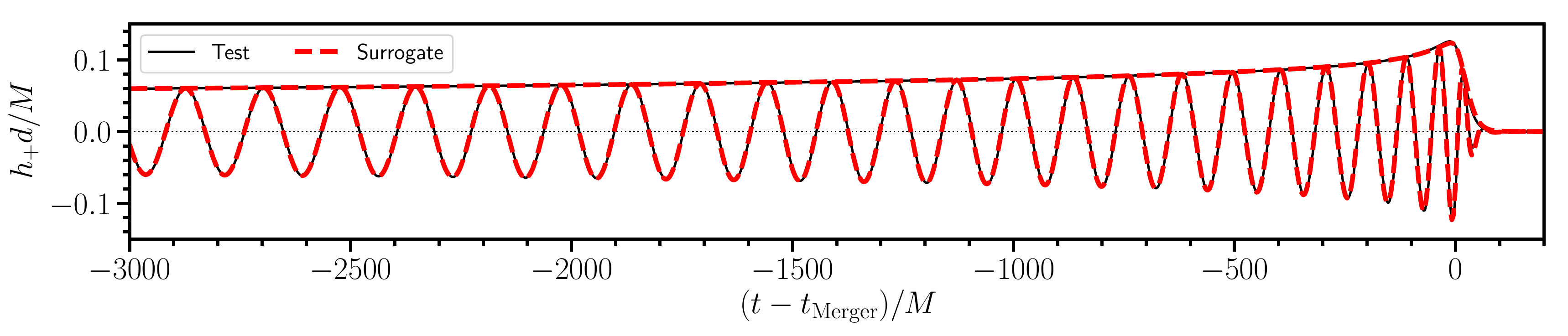}
\caption{Inverse Fourier transformed surrogate (red dashed) for the parameters $\{q, \chi_1, \chi_2, \Lambda_1, \Lambda_2\} = \{0.61, -0.23, -0.33, 180, 4390\}$ that have the largest mismatch with the test-set EOB waveforms (black). The $+$ polarizations and amplitudes are shown. The two waveforms are aligned in the time domain, before the start of the plot, in the interval $(t-t_{\rm Merger})/M \in [-10^7, -10^6]$.}
\label{fig:maxmismatchtd}
\end{figure*}

Finally, to demonstrate that we can correctly recover the behavior of the original, time-domain EOB waveform, we inverse Fourier transform the surrogate and compare it to the time-domain waveform. In Fig.~\ref{fig:maxmismatchtd} we compare the final surrogate to the test set waveform that has the largest mismatch. We align the two waveforms in time and phase by maximizing the overlap at early times in the interval $(t-t_{\rm merger})/M \in [-10^7, -10^6]$ using the method of~\cite{ReadMarkakisShibata2009}. The two waveforms remain nearly identical up to merger, far from the interval where they were aligned.

\subsection{Timing}

The surrogate evaluation time is dominated by two parts. The first part is the time needed to evaluate the residuals $\{\mathcal{I}_{\rm GPR}[\Delta\ln A_j](\bx)\}$ and $\{\mathcal{I}_{\rm GPR}[\Delta\Phi_j](\bx)\}$ at each of the $N_A+N_\Phi$ interpolating nodes using GPR. Although optimizing the hyperparameters for GPR scales with the number of training set samples $N$ as $\mathcal{O}(N^3)$, the evaluation time of a stored GPR scales as $\mathcal{O}(N)$ (see Eq.~\eqref{eq:mean}). The evaluation time for all the residuals therefore has a cost of $\mathcal{O}(N(N_A+N_\Phi))$. The second part is the time needed to resample the final surrogate (Eqs.~\eqref{eq:amp_sur}--~\eqref{eq:hcross_sur}) at uniformly spaced frequency samples in physical units beginning at a starting frequency $f_{\rm min}$. 

\begin{figure}[htb]
\centering
\includegraphics[width=0.49\textwidth]{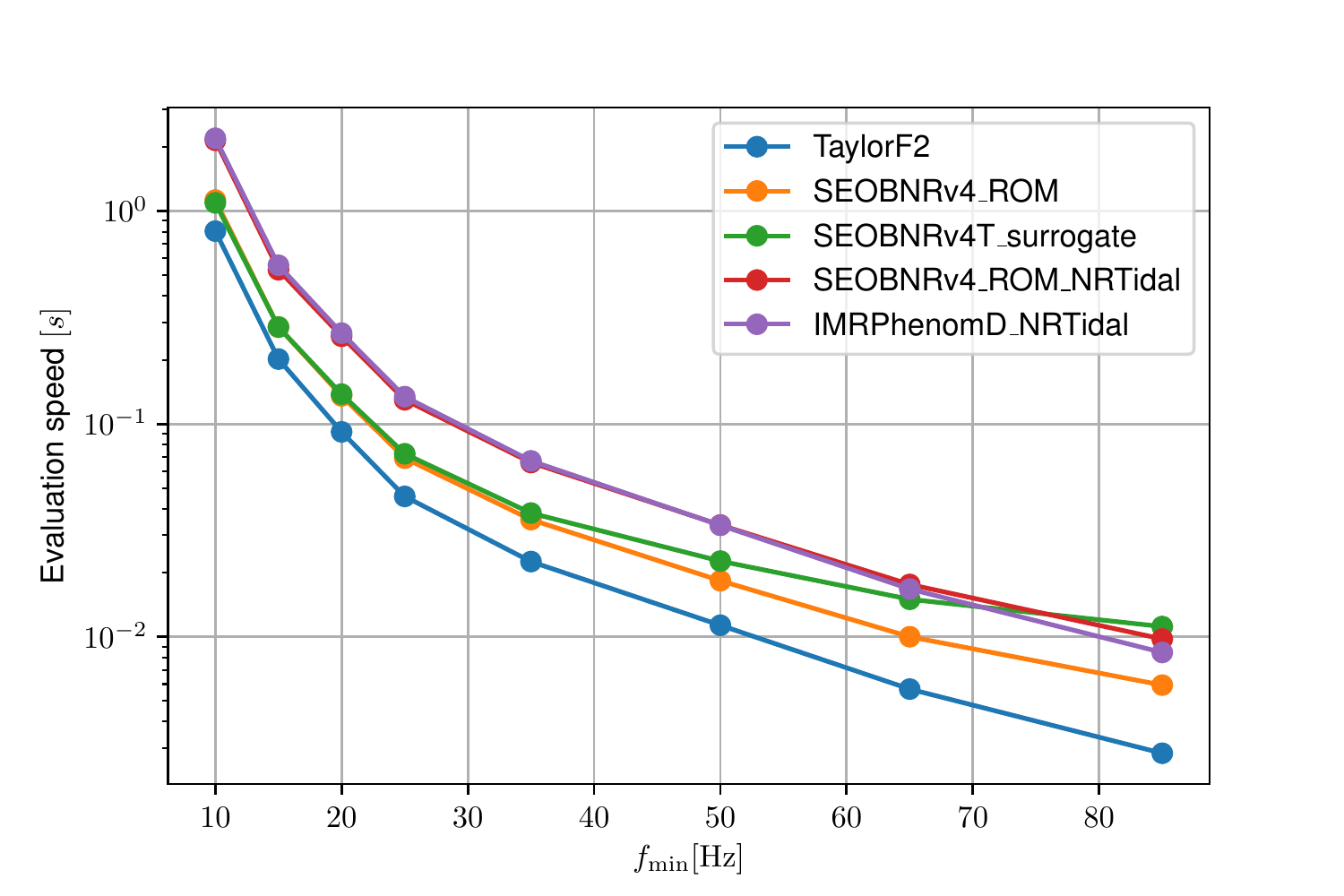}
\includegraphics[width=0.49\textwidth]{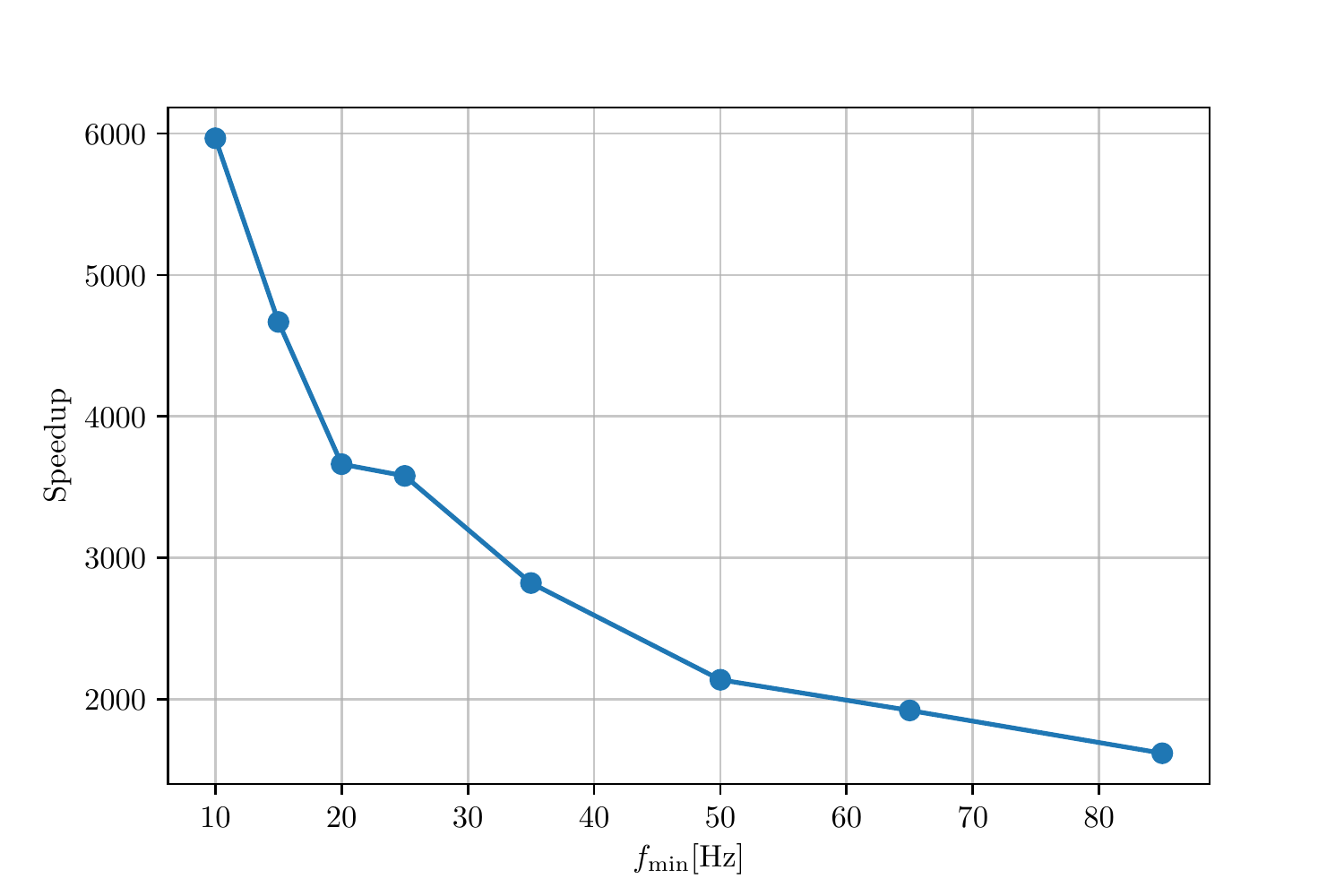}
\caption{Waveform evaluation time as a function of the waveform starting frequency $f_{\rm min}$. The time-domain \texttt{SEOBNRv4T} waveform was sampled at 4096~Hz then Fourier transformed. The frequency-domain waveforms used the same frequency samples as the Fourier transformed time-domain waveforms. Top panel: Evaluation times of the frequency-domain waveforms. Bottom panel: Speedup of \texttt{SEOBNRv4T\_surrogate} relative to the original \texttt{SEOBNRv4T}.}
\label{fig:timing}
\end{figure}

In Fig.~\ref{fig:timing}, we show the surrogate evaluation time as a function of the starting frequency $f_{\rm min}$ for an equal mass $1.4M_\odot$--$1.4M_\odot$ binary evaluated on a 3.5~GHz Intel Xeon processor. The waveform was sampled with a uniform frequency spacing corresponding to a sampling rate of 4096~Hz. At large starting frequencies the evaluation time is limited by the $\sim 0.01$~s needed to calculate the GPR fits at each frequency node. As $f_{\rm min}$ decreases, the waveform becomes longer and a smaller frequency spacing is needed. The evaluation time is then dominated by the spline interpolation needed to resample the waveform. We also compare the evaluation time to three BNS waveform models used recently in the analysis of GW170817 (\texttt{TaylorF2}, \texttt{SEOBNRv4\_ROM\_NRTidal}, \texttt{IMRPhenomD\_NRTidal})~\cite{BNSPE} and the BBH model \texttt{SEOBNRv4\_ROM}. Only the analytic \texttt{TaylorF2} model is noticeably faster. In the bottom panel, we compare the evaluation time to the original time-domain \texttt{SEOBNRv4T} waveform model and find it is $\sim 1500$--6000 times faster.

\section{Parameter estimation}
\label{sec:pe}

As a final test of the surrogate, we perform several parameter estimation runs where we inject the original, time-domain \texttt{SEOBNRv4T} waveform and recover the injected parameters with the \texttt{SEOBNRv4T\_surrogate} template. We use the design sensitivity PSDs for the two advanced LIGO detectors and the advanced Virgo detector~\cite{Aasi:2013wya}. We choose as our waveform parameters $m_1=m_2=1.4M_\odot$, $\chi_1=\chi_2=0.1$, and $\Lambda_1=\Lambda_2=1286$ corresponding to the EOS named MS1b in~\cite{ReadLackey2009}. The sky location and inclination angle are fixed at an arbitrary value, and the distances are chosen such that the optimal network signal-to-noise ratios (SNRs) are either 30, 60, or 120. We inject the waveforms into zero noise data so that the recovered parameters do not depend on the specific noise realization~\cite{NissankeHolzHughes2010}.

As discussed in more detail in~\cite{BNSPE}, the posterior $p(\vec\vartheta | d)$ for the parameters $\vec\vartheta$ given the detector data $\vec d$ is given by Bayes' theorem
\begin{equation}
p(\vec\vartheta | \vec d) \propto p(\vec\vartheta) \mathcal{L}(\vec d | \vec\vartheta),
\end{equation}
where $p(\vec\vartheta)$ is the prior and $\mathcal{L}(\vec d | \vec\vartheta)$ is the likelihood. 
For the prior we choose the sky position and orientation to be uniform on the unit sphere; the distance to be uniform in co-moving volume; the masses, spins, and tidal parameters to be uniform in the ranges $m_{1, 2} \in [0.7, 2]M_\odot$, $\chi_{1,2} \in [-0.5, 0.5]$ and $\Lambda_{1,2} \in [0, 5000]$; and the mass ratio is restricted to the range $q\in[0.5, 1]$. 
As was done in~\cite{BNSPE}, we sample the data and waveform at 4096~Hz such that the Nyquist frequency is 2048~Hz, and integrate the likelihood function in the interval $[20, 2048]$~Hz. We sample this posterior with the parallel-tempered MCMC code in \texttt{LALInference}~\cite{lal} which requires $\sim 10^7$ iterations to produce 10,000--20,000 statistically independent samples. 

We note that the \texttt{SEOBNRv4T} waveform model still has some signal above 2048~Hz (see Fig.~\ref{fig:hoff}). We therefore also performed runs with a sampling frequency of 8192~Hz and integrated the likelihood function in the interval $[20, 4096]$~Hz. However, this did not have a noticeable effect on the recovered posteriors.

\begin{figure}[htb]
\centering
\includegraphics[width=0.49\textwidth]{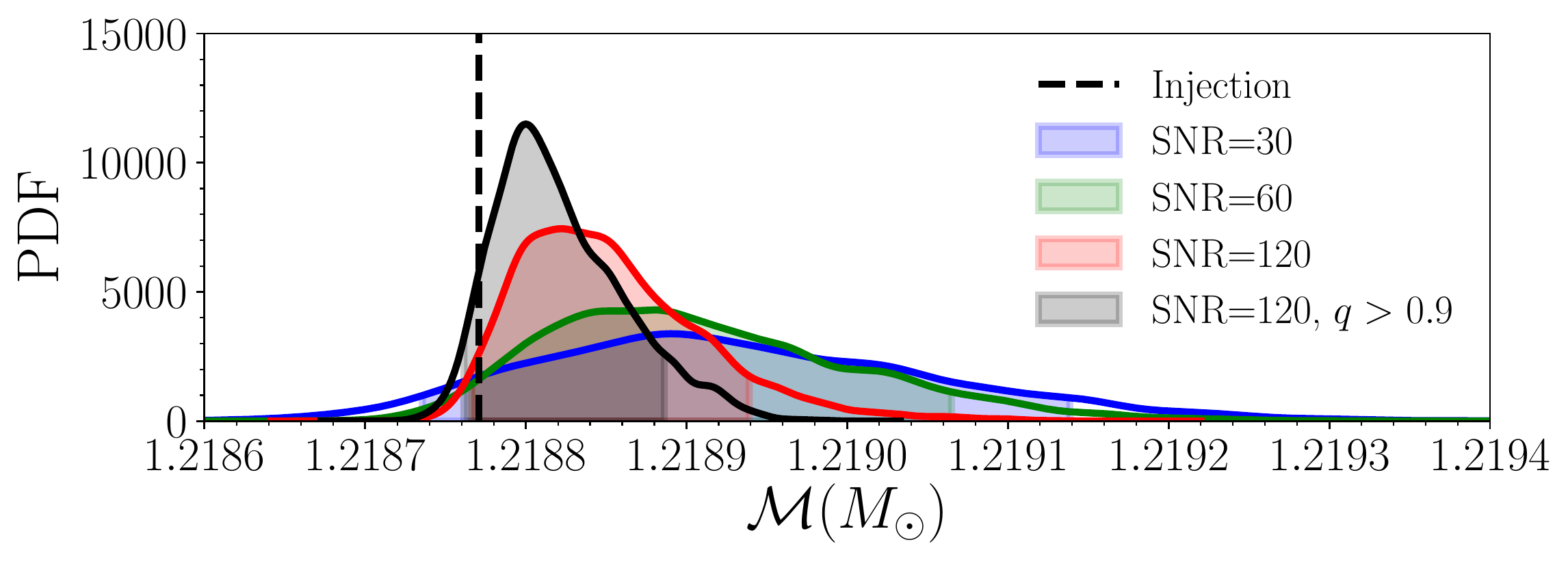}\\
\includegraphics[width=0.49\textwidth]{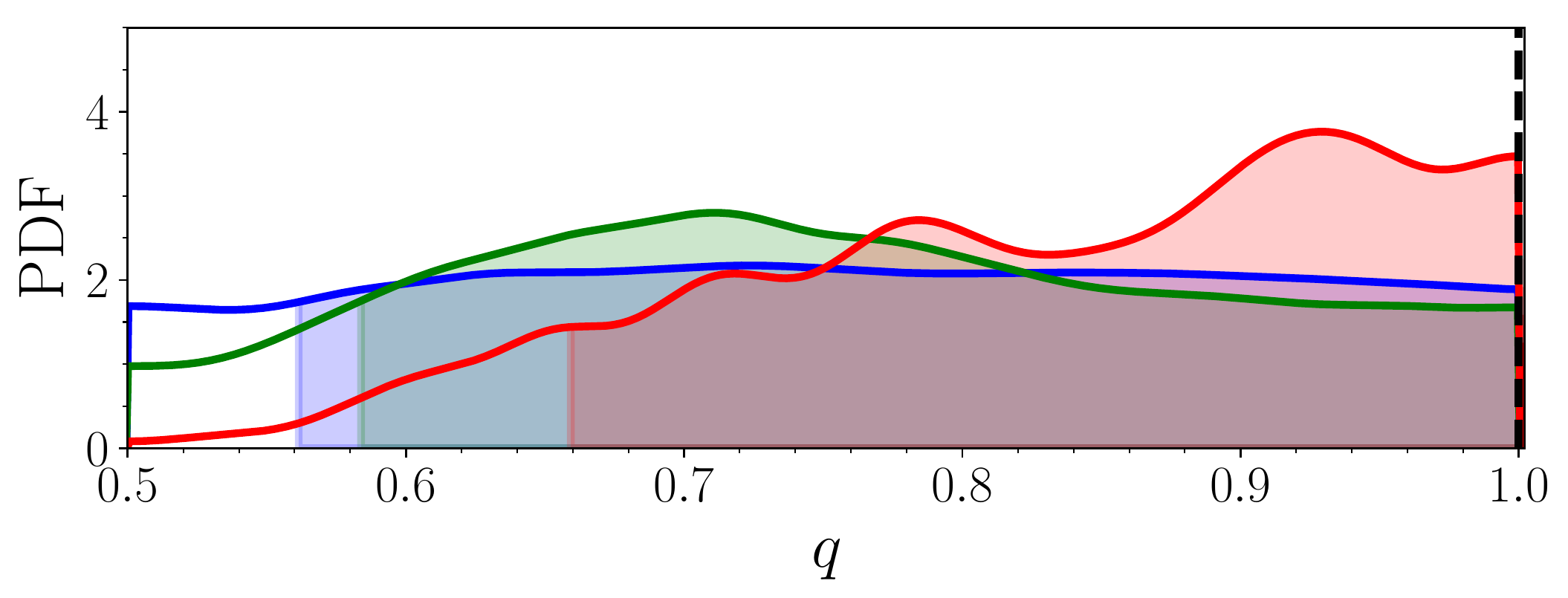}\\
\includegraphics[width=0.49\textwidth]{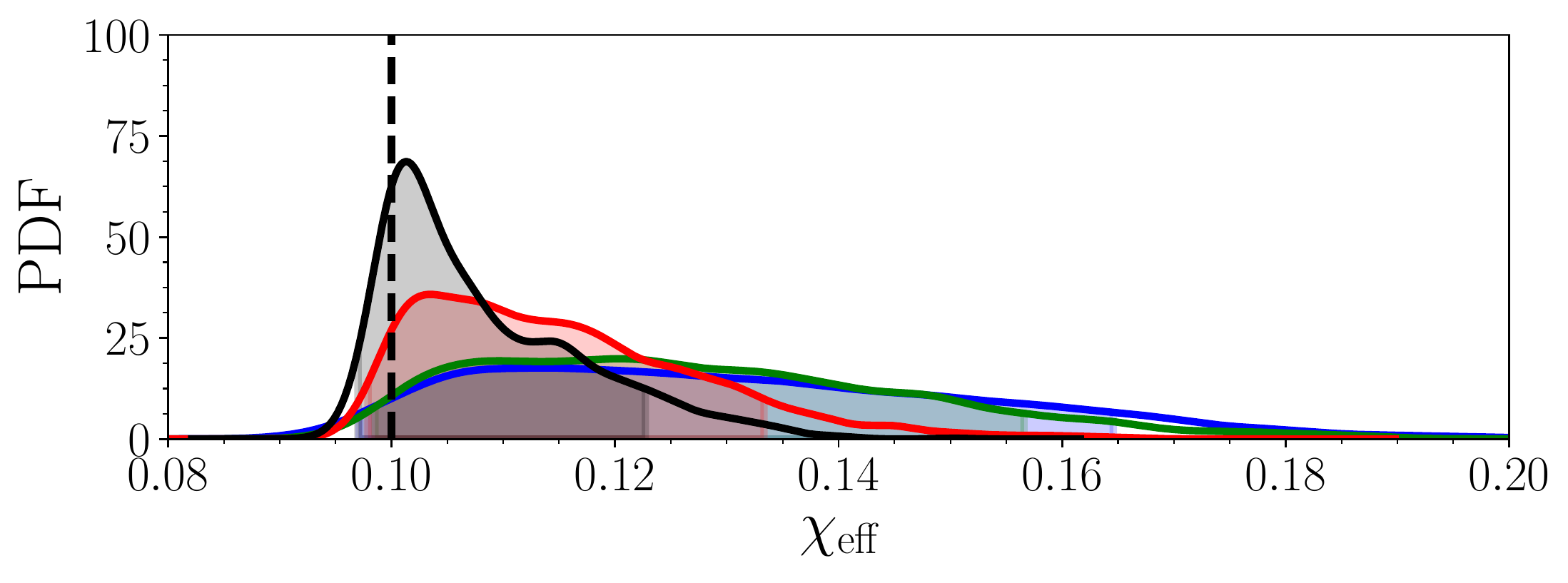}\\
\includegraphics[width=0.49\textwidth]{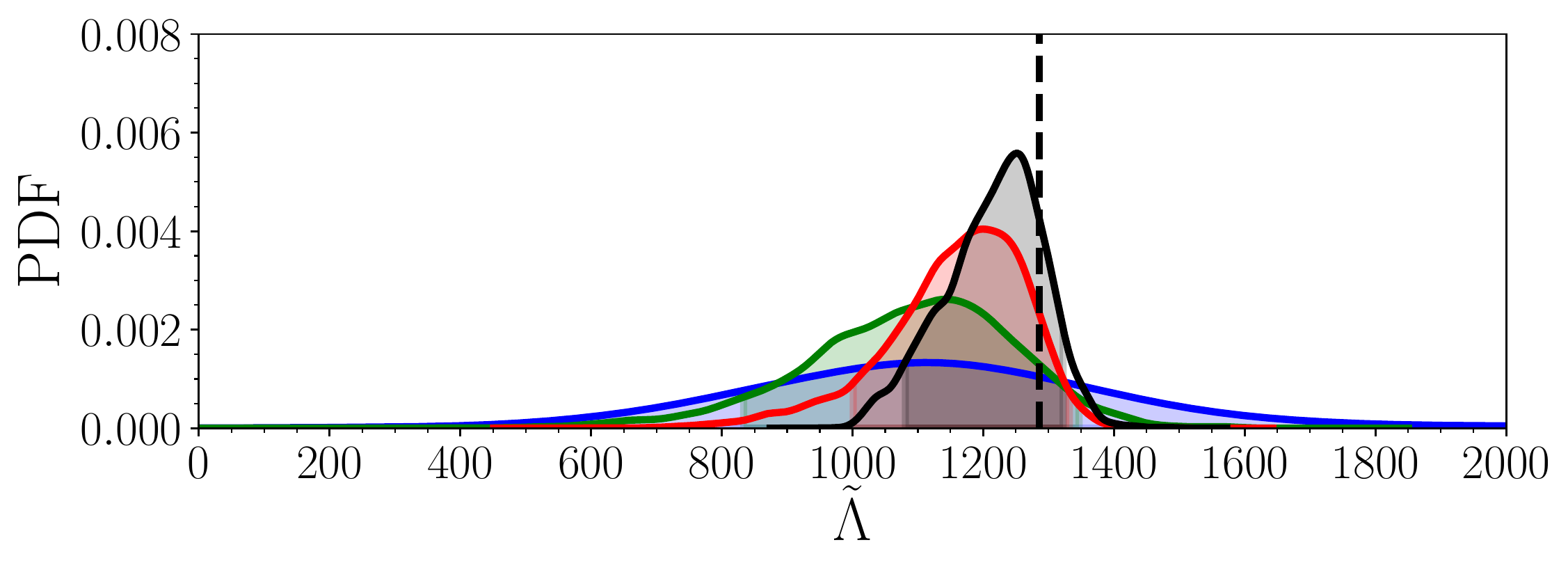}
\caption{Marginalized 1-d PDFs for waveform injections with zero noise. The parameters are identical except for the distances which are chosen such that the network SNRs are 30 (blue), 60 (green), or 120 (red). The black PDFs only include samples for the SNR=120 injection that satisfy $q>0.9$. The shaded regions give the 90\% credible intervals. The dashed black line represents the injected parameter value.}
\label{fig:1dpe}
\end{figure}

\begin{figure}[htb]
\centering
\includegraphics[width=0.49\textwidth]{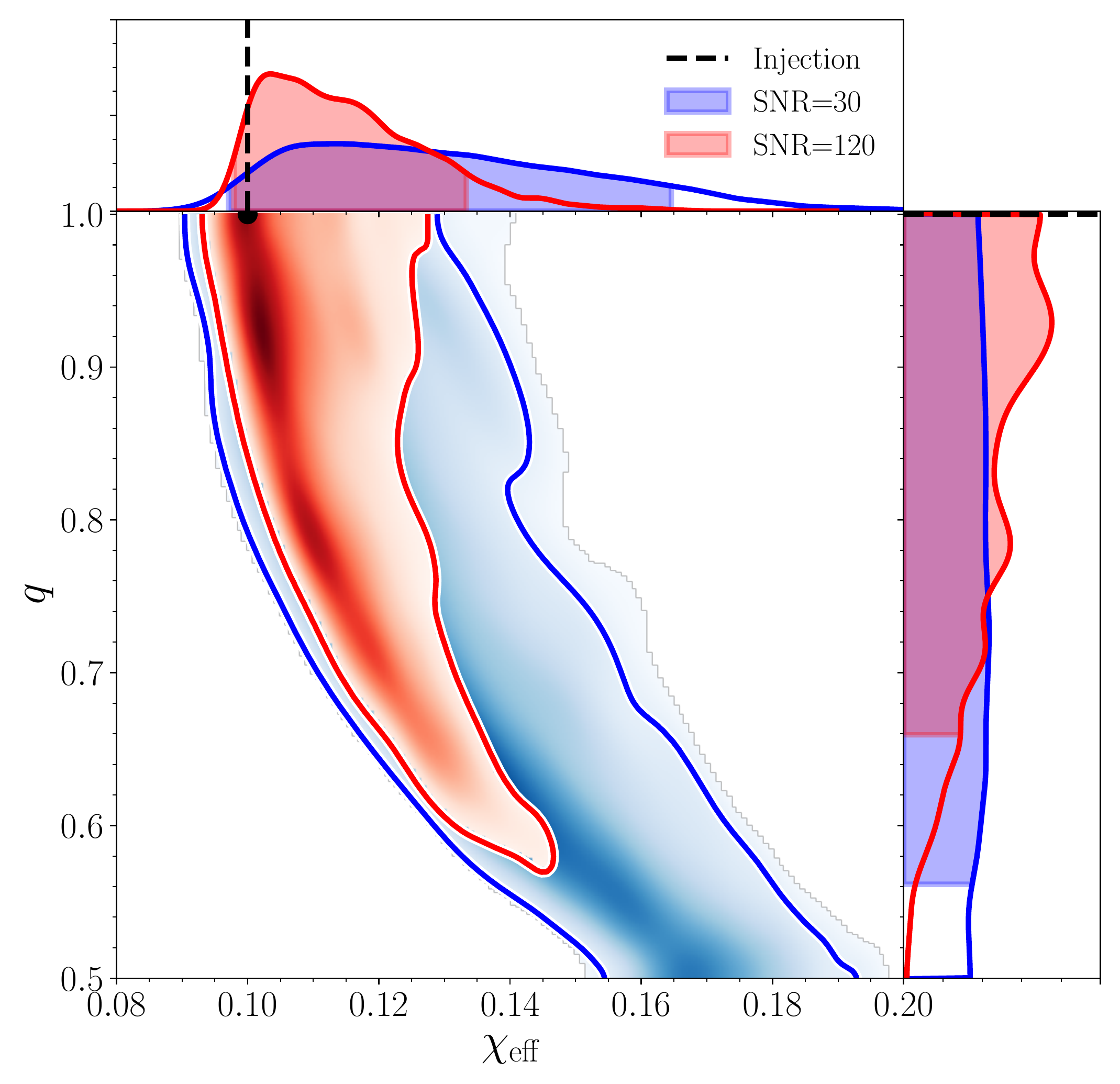}
\caption{Marginalized 1-d and 2-d PDFs for the parameters $\chi_{\rm eff}$ and $q$ for the injections with SNRs of 30 (blue) and 120 (red). The shaded regions for the 1-d PDFs give the 90\% credible interval. The contours for the 2-d PDFs give the 90\% credible regions. The dashed black lines and circle represent the injected parameter value.}
\label{fig:2dpe}
\end{figure}

We show marginalized 1-dimensional posteriors for four key parameters in Fig.~\ref{fig:1dpe}. These are the chirp mass $\mathcal{M} = (m_1 m_2)^{3/5}/(m_1+m_2)^{1/5}$, the mass ratio $q=m_2/m_1$, the effective spin $\chi_{\rm eff} = (m_1\chi_1+m_2\chi_2)/(m_1+m_2)$ and the effective tidal deformability parameter $\tilde\Lambda = (16/13) [(1+12q)\Lambda_1 + (12+q)q^4\Lambda_2]/(1+q)^5$. As the SNR increases, the peaks of the marginalized 1-d posteriors become more closely aligned with the injected value. However, the distributions are noticeably asymmetric and are not centered on the injected values. 
This results mainly because prior boundaries on some parameters, such as the mass ratio $q$, shift the marginalized PDFs for correlated parameters.
This effect is strongest for the interaction between $\chi_{\rm eff}$ and $q$ (Fig.~\ref{fig:2dpe}), where the injected value $q=1$ is at the boundary of the prior. Although the injected parameter $(\chi_{\rm eff}, q) = (0.1, 1)$ lies on the ridge of maximum density in the marginalized 2-d PDF, the marginalized 1-d PDF for $\chi_{\rm eff}$ is noticeably offset. To counteract this effect in Fig.~\ref{fig:1dpe}, we also take a thin cross section containing the injected value of $q=1$ by cutting samples outside the interval $q \in [0.9, 1]$. This causes the peaks of the other 1-d distributions to be much closer to the injected values. We consider the difference between these peaks and the injected values to be a conservative bound on the size of the systematic errors due to errors in the surrogate model.

\section{Discussion and future work}
\label{sec:discussion}

We have constructed a fast, frequency-domain surrogate of one of the most accurate BNS waveform models to date. This aligned-spin model, \texttt{SEOBNRv4T}, incorporates the tidally induced $\ell=2$ and 3 multipole moments as well as the effect of dynamical tides as the excitation approaches the $\ell=2$ and 3 $f$-mode frequencies. We have achieved mismatches of no more than $4.5 \times 10^{-4}$ and phase errors of $\lesssim 1$~rad up to the merger frequency. These are sufficient to not bias results in any of the parameters.

The evaluation time has a flat cost of $\sim 0.01$~s to perform the GPR interpolation at each node. The rest of the time is spent resampling the waveform with spline interpolation. For a starting frequency of 20~Hz, this takes 0.13~s when the waveform is matched with data uniformly sampled at 4096Hz. For current parameter estimation codes, this is sufficient. However, one could further improve run times using reduced order quadrature~\cite{Antil2013, CanizaresFieldGair2013, CanizaresFieldGair2015, Smith:2016qas} which requires frequency-domain waveforms, multi-band waveform interpolation~\cite{VinciguerraVeitchMandel2017}, or relative binning~\cite{Zackay:2018qdy}. Finally, we note that the production MCMC sampler used for the GW170817 analysis has an autocorrelation length of $\mathcal{O}(10^3)$ for aligned-spin BNS models, so significant improvements to the parameter estimation runtime can be made through better samplers. 

The hierarchical method presented here, where we begin with the analytic TaylorF2 reference model then make a surrogate of the residual, can be used to further improve the waveform model. For example, one could build a surrogate for numerical BNS simulations using \texttt{SEOBNRv4T\_surrogate} as the base model, then constructing a surrogate of the residual. Such a model would have the accuracy of EOB below $\sim 400$~Hz and the accuracy of numerical simulations for the last several cycles before merger. Current state-of-the-art NR simulations have phase errors of several tenths of a radian over the last $\sim 20$ gravitational-wave cycles~\cite{DietrichHinderer2017, KiuchiKawaguchiKyutoku2017}. Using GPR and uncertainty sampling discussed above, one could optimally choose the waveform parameters for the numerical simulations, and run them in parallel to build the training set. If the difference between \texttt{SEOBNRv4T} and numerical BNS simulations is small, one would not need a high fractional accuracy for a surrogate of the difference, and 10--100 waveforms may be sufficient.

This model notably does not include precession for non-aligned spins. Although none of the EOB models currently available include both tidal effects and precession, there are ways to rotate this waveform model to a precessing frame and approximately incorporate precession. For example, Chatziioannou {\it et al}.\ have analytically solved the 2PN-accurate precession equations for generic spins~\cite{ChatziioannouKleinCornish2017a, ChatziioannouKleinCornish2017b}, with the exception of transitional precession which is unlikely for low mass-ratio and spin BNS systems. Using shifted uniform asymptotics~\cite{Klein:2014bua}, they can also analytically Fourier-transform the solution. Importantly, one is free to specify the frequency-domain amplitude and phase evolution, such as the \texttt{SEOBNRv4T\_surrogate} here, for the waveform in the co-precessing frame. This approach would provide a fast, accurate model for BNS systems with tides and generic spins. We note, however, that including precession did not noticeably affect the parameter estimation results for GW170817~\cite{BNSPE}.

\begin{acknowledgments}

The authors sincerely thank Justin Vines and Tanja Hinderer for deriving the quadrupole-monopole corrections needed in the model. The authors are grateful to Zoheyr Doctor, Bhooshan Gadre, and Prayush Kumar for reviewing and testing the implementation of \texttt{SEOBNRv4T\_surrogate} in \texttt{LAL}. BL also thanks the participants of the Bayesian Methods in Nuclear Physics workshop at the University of Washington for helpful advice at the start of this work.

\end{acknowledgments}

\appendix
\section{TaylorF2 reference waveform}
\label{sec:taylorf2}

The explicit expressions for the amplitude and phase of the TaylorF2 waveform are as follows. For the amplitude we use the 1PN correction to the leading order waveform,
\begin{equation}
A_{\rm F2} = -\sqrt{\frac{5\pi\eta}{24}} v^{-7/2} \left[ 1 + \left(-\frac{323}{224} + \frac{451\eta}{168} \right)v^{2} \right],
\end{equation}
where $v=(\pi M f)^{1/3}$ is the standard PN parameter. The sign is set following the conventions of \texttt{LAL}~\cite{lal}. The phase has the schematic form
\begin{align}
\Phi_{\rm F2} =& -2\pi f t_{0} + \phi_{0} + \frac{\pi}{4} - \frac{3}{128\eta}v^{-5} \left[\Phi_{\rm F2}^{\rm PP}(\eta) \right. \nonumber\\
	& \left. \qquad + \Phi_{\rm F2}^{\rm Spin} (\eta, \chi_1, \chi_2) + \Phi_{\rm F2}^{\rm Tidal}(\eta, \Lambda_1, \Lambda_2)  \right] \,,
\end{align}
where $t_{0}$ and $\phi_{0}$ are constants related to the freedom of choosing the time and phase of coalescence. We separated point particle terms $\Phi_{\rm F2}^{\rm PP}(\eta)$, spin terms $\Phi_{\rm F2}^{\rm Spin}(\eta, \chi_{1}, \chi_{2})$ and tidal terms $\Phi_{\rm F2}^{\rm Tidal}(\eta, \Lambda_1, \Lambda_2)$. We include point particle terms to the highest know order of 3.5PN (see e.g.~\cite{Blanchet2014Review}).

The spin terms can be decomposed into spin-orbit and spin-spin terms (ignoring cubic-in-spin contributions at 3.5PN) as $\Phi_{\rm F2}^{\rm Spin} = \Phi_{\rm F2}^{\rm SO}  + \Phi_{\rm F2}^{\rm SS} $. We include spin-orbit terms up to 3.5PN~\cite{Blanchet2014Review}. For spin-square terms, as the \texttt{SEOBNRv4T} waveform does not include yet 3PN spin-spin effects (see Sec.~\ref{subsec:QM}), we stop at  the leading 2PN in the \texttt{TaylorF2} waveform and ignore the known 3PN terms. The main reason for doing this is because including only common physical effects in both \texttt{SEOBNRv4T} and \texttt{TaylorF2} makes the residual $\Delta\Phi(Mf; \bx)$ smaller and have less variation, making it easier to fit at a given error requirement.

The tidal terms $\Phi_{\rm F2}^{\rm Tidal}(\eta, \Lambda_1, \Lambda_2)$ are known to 6PN order~\cite{VinesFlanaganHinderer2011}. These tidal contributions to the phase are exactly as used in the \texttt{LAL} waveform \texttt{TaylorF2}.

For completeness, we give below explicit expressions of the PN coefficients entering the phasing, using the notations $\delta \equiv (m_{1} - m_{2})/(m_{1} + m_{2})$ for the mass difference, and $\chi_{s,a} = (\chi_{1} \pm \chi_{2})/2$, $\kappa_{s,a} = (\kappa_{1} \pm \kappa_{2})/2$ for the symmetrized and antisymmetrized combinations of $\chi_{A}$, $\kappa_{A}$.

The point particle terms coefficients up to 3.5PN can be found e.g. in Eq.~(3.18) of Ref.~\cite{BuonannoIyerOchsner2009}, and have the structure
\be
	\Phi_{\rm F2}^{\rm PP}(\eta) = \sum_{k=0}^{7} a_{k} v^{k} + \mathcal{O}(v^{8}) \,.
\ee
The non-zero coefficients read
\begin{align}
	a_{0} &= 1 \nonumber\\
	a_{2} &= \frac{55 \eta }{9}+\frac{3715}{756} \nonumber\\
	a_{3} &= -16 \pi \nonumber\\
	a_{4} &= \frac{3085 \eta ^2}{72}+\frac{27145 \eta }{504}+\frac{15293365}{508032} \nonumber\\
	a_{5} &= -\frac{65 \pi  \eta }{9}-\frac{65}{3} \pi  \eta   \ln v +\frac{38645}{252} \pi   \ln v +\frac{38645 \pi }{756} \nonumber\\
	a_{6} &= -\frac{127825 \eta ^3}{1296}+\frac{76055 \eta ^2}{1728}+\frac{2255 \pi ^2 \eta }{12} \nonumber\\ 
	& -\frac{15737765635 \eta }{3048192}-\frac{6848  \ln v }{21}-\frac{640 \pi ^2}{3}-\frac{6848 \gamma_{E} }{21} \nonumber\\ 
	& +\frac{11583231236531}{4694215680}-\frac{13696 \log (2)}{21} \nonumber\\
	a_{7} &= -\frac{74045 \pi  \eta ^2}{756}+\frac{378515 \pi  \eta }{1512}+\frac{77096675 \pi }{254016} \,,
\end{align}
where $\gamma_{E}$ is Euler's constant.

The spin-orbit and spin-spin corrections have the structure
\begin{align}
	\Phi_{\rm F2}^{\rm SO}(\eta, \chi_{1}, \chi_{2}) &= \sum_{k=3}^{7} b_{k} v^{k} + \mathcal{O}(v^{8})\nonumber\\
	\Phi_{\rm F2}^{\rm SS}(\eta, \chi_{1}, \chi_{2}) &= c_{4} v^{4} + c_{6} v^{6} + \mathcal{O}(v^{7}) \,.
\end{align}
\begin{widetext}
The non-zero spin-orbit coefficients are
\begin{align}
	b_{3} &= \frac{113 \delta   \chi_a }{3}+\left(\frac{113}{3}-\frac{76 \eta }{3}\right)  \chi_s \nonumber\\
	b_{5} &=  \chi_s  \left(\frac{340 \eta ^2}{9}+\frac{24260 \eta }{81}+\frac{340}{3} \eta ^2  \ln v +\frac{24260}{27} \eta   \ln v -\frac{732985  \ln v }{756}-\frac{732985}{2268}\right) \nonumber\\
	& \quad + \chi_a  \left(-\frac{140 \delta  \eta }{9}-\frac{732985 \delta }{2268}-\frac{140}{3} \delta  \eta   \ln v -\frac{732985}{756} \delta   \ln v \right)  \nonumber\\
	b_{6} &= \frac{2270 \pi  \delta   \chi_a }{3}+\left(\frac{2270 \pi }{3}-520 \pi  \eta \right)  \chi_s \nonumber\\
	b_{7} &=  \chi_a  \left(-\frac{1985 \delta  \eta ^2}{48}+\frac{26804935 \delta  \eta }{6048}-\frac{25150083775 \delta }{3048192}\right)+\left(\frac{5345 \eta ^3}{36}-\frac{1042165 \eta ^2}{3024}+\frac{10566655595 \eta }{762048}-\frac{25150083775}{3048192}\right)  \chi_s \,,
\end{align}
while the leading-order quadratic-in-spin term reads
\begin{align}
	c_{4} &= \chi_s ^2 \left(-50 \delta   \kappa_a +100  \kappa_s  \eta -50  \kappa_s -\frac{195 \eta }{2}-\frac{5}{8}\right) + \chi_a   \chi_s  \left(-100 \delta   \kappa_s -\frac{5 \delta }{4}+200  \kappa_a  \eta -100  \kappa_a \right) \nonumber\\
	& + \chi_a ^2 \left(-50 \delta   \kappa_a +100  \kappa_s  \eta -50  \kappa_s +100 \eta -\frac{5}{8}\right) \,.
\end{align}
As explained above, to be consistent with the fact that this PN information has not yet been incorporated in \texttt{SEOBNRv4T}, we do not include the next-to-leading 3PN term, which for reference is given by~\cite{Bohe:2015ana}
\begin{align}
	c_{6} &=  \chi_s ^2 \left(-\frac{1495 \delta   \kappa_a  \eta }{6}+\frac{26015 \delta   \kappa_a }{28}-240  \kappa_s  \eta ^2-\frac{44255  \kappa_s  \eta }{21}+\frac{26015  \kappa_s }{28}+\frac{3415 \eta ^2}{9}+\frac{829705 \eta }{504}-\frac{1344475}{2016}\right) \nonumber\\
	& + \chi_a   \chi_s  \left(-\frac{1495 \delta   \kappa_s  \eta }{3}+\frac{26015 \delta   \kappa_s }{14}+\frac{745 \delta  \eta }{18}-\frac{1344475 \delta }{1008}-480  \kappa_a  \eta ^2-\frac{88510  \kappa_a  \eta }{21}+\frac{26015  \kappa_a }{14}\right) \nonumber\\
	& + \chi_a ^2 \left(-\frac{1495 \delta   \kappa_a  \eta }{6}+\frac{26015 \delta   \kappa_a }{28}-240  \kappa_s  \eta ^2-\frac{44255  \kappa_s  \eta }{21}+\frac{26015  \kappa_s }{28}-240 \eta ^2+\frac{267815 \eta }{252}-\frac{1344475}{2016}\right)\,.
\end{align}
\end{widetext}

Finally, the tidal contributions to the phasing take the form~\cite{VinesFlanaganHinderer2011}
\begin{equation}
	\Phi_{\rm F2}^{\rm Tidal}(\eta, \Lambda_{1}, \Lambda_{2}) = d_{10} v^{10} + d_{12} v^{12} + \mathcal{O}(v^{13}) \,.
\end{equation}
Introducing convenient mass-weighted combinations of the tidal parameters, expressed using $X_{1} = m_{1}/(m_{1} + m_{2})$ and $X_{2} = m_{2}/(m_{1} + m_{2})$,
\begin{align}
	\tilde\Lambda &= \frac{16}{13} \left[(12 - 11 X_1)X_1^4\Lambda_1 + (12 - 11 X_2)X_2^4\Lambda_2 \right] \nonumber\\
	\delta\tilde\Lambda =& \frac{1}{1319} \left[(-11005 + 14014 X_1 - 1690 X_{1}^{2})X_1^4\Lambda_1 \right. \nonumber\\ 
	& \left. \quad\quad + (11005 - 14014 X_2 + 1690 X_{2}^{2}) X_2^4\Lambda_2 \right] \,,
\end{align}
the coefficients are given by
\begin{align}
	d_{10} &= -\frac{39}{2}\tilde\Lambda \nonumber\\
	d_{12} &= -\frac{3115}{64}\tilde\Lambda + \frac{6595}{364}(X_1-X_2)\delta\tilde\Lambda \,.
\end{align}

\bibliography{paper,refs}

\end{document}